# The Impact of Large Language Models on Open-Source Innovation: Evidence from GitHub Copilot[*]

Doron Yeverechyahu
Coller School of Management, Tel Aviv University

Raveesh Mayya
Stern School of Business, New York University

Gal Oestreicher-Singer
Coller School of Management, Tel Aviv University

## Abstract

Large Language Models (LLMs) are reshaping knowledge work, yet their impact on voluntary, self-guided open innovation forums (settings where contributors choose their own tasks without managerial direction) may differ fundamentally from effects observed in organizational settings. We study this question in open-source software development, where individuals' contributions collectively drive innovation at a community level. Unlike product innovation, where typologies for classifying innovation are well established, knowledge work in open-source settings calls for a distinction grounded in the cognitive demand a task places on the contributor. Burgeoning literature distinguishes *substantive* contributions, which require creative problem formulation to introduce new functionality, from *incremental* contributions, which draw on comprehension of existing code to maintain and refine it. We exploit a natural experiment around GitHub Copilot's launch in October 2021, where Copilot supported languages like Python while not supporting R for business reasons, creating an exogenous partition between otherwise comparable ecosystems. Using three complementary and independent identification strategies and two classification approaches, we find that Copilot availability increases the volume of open-source contributions by 28 to 40 percent. The increase in incremental contributions is significantly larger than the increase in substantive contributions across all specifications. This disparity is more pronounced in projects with higher activity levels and widens following a model upgrade: LLMs function more effectively when existing context (prior code, documentation, issue discussions) helps define the problem and constrain solutions, tilting collaborative innovation toward the exploitation of established codebases rather than the exploration of new functionality. This paper provides one of the rare instances of causal field evidence on LLM effects, given the speed at which generative AI has exploded across the knowledge economy. Our findings highlight the need for open-source platforms and organizations that encourage open-source contributions to introduce deliberate incentives for substantive feature development.

---

[*] All authors contributed equally to this work. Author names are listed in reverse order of academic seniority.
G. Oestreicher-Singer gratefully acknowledges the funding from the European Union's Horizon 2020 research and innovation program (#759540). R.Mayya gratefully acknowledges the funding from the Center for Global Economy and Business (CGEB) at NYU Stern.

## 1. Introduction

Rapid advances in Large Language Model (LLM) technologies are promising to transform various aspects of economic activities. Burgeoning research has explored how LLMs can affect the future of work and labor markets (Brynjolfsson et al. 2018; Eloundou et al. 2024), productivity and creativity (Boussioux et al. 2024; Brynjolfsson et al. 2025; Chen and Chan 2024; Zhou and Lee 2024; Doshi and Hauser 2024), and knowledge dissemination (Burtch et al. 2024; Horton 2023; Quinn and Gutt 2025; Ma et al. 2025).

How LLMs reshape collaborative innovation in voluntary, platform-mediated knowledge ecosystems remains an open empirical question. Such ecosystems differ from firm-bounded settings because contributors self-select into projects and tasks rather than receiving direction from a manager (Felin and Zenger 2014). Open-source software development is the leading example of such a setting: contributions[1] are voluntary, modular, and digitally native, and coordination happens through commits, pull requests, and issue threads on shared platforms (von Hippel and von Krogh 2003; Lin and Maruping 2022). Unlike earlier open-source automation tools such as continuous-integration pipelines or dependency-update bots, which automate task execution, LLMs operate through natural language, helping contributors reason through and complete their code. They engage with the cognitive content of individual contributions (Agrawal et al. 2018), arguably reshaping the composition of innovation emerging from such contributions.

Because LLMs assist not just with executing code but with reasoning through it, the classical typologies developed for firm-bounded product innovation (Henderson and Clark 1990) are not well suited to characterize how their effects might play out in voluntary collaborative settings. We need typologies that anchor different ends of a cognitive spectrum, differing in knowledge search strategies and resultant competence outcomes (March 1991; Gatignon et al. 2002). We build on recent open-source innovation research (Huang et al. 2026) and label these as *substantive* innovation and *incremental* innovation. *Substantive* innovation involves a search process focused on seeking knowledge beyond the project's current domains and acquiring new competence. In collaborative environments like open source, substantive innovation requires "question ingenuity": the cognitive ability to identify entirely new problem spaces worth exploring before trying to generate solutions. Contributors performing substantive work must conceptualize features without predetermined specifications, demanding substantial human input to conceptualize and execute approaches that lack precedent within the project context. *Incremental* innovation involves a search process centered on deepening understanding within existing knowledge domains and strengthening existing competence. Incremental innovation in open source starts with understanding existing code, often written by others, before making targeted improvements that navigate complex codebase interdependencies (Medappa and Srivastava 2019). This can happen with limited human input because the existing work itself forms part of the input along with the type of improvement needed. Figure 1 illustrates this in a data visualization package context: Panel A shows an incremental contribution where readability for black-and-white printing is improved in an existing visualization, and Panel B shows a substantive contribution where confidence interval display is added to the package.

[1] The open-source literature distinguishes a contribution, an individual act such as a commit, fix, or feature addition, from innovation, the higher-order construct that emerges when many contributions accumulate (von Hippel and von Krogh 2003; Dahlander and Magnusson 2008). We follow this convention throughout.

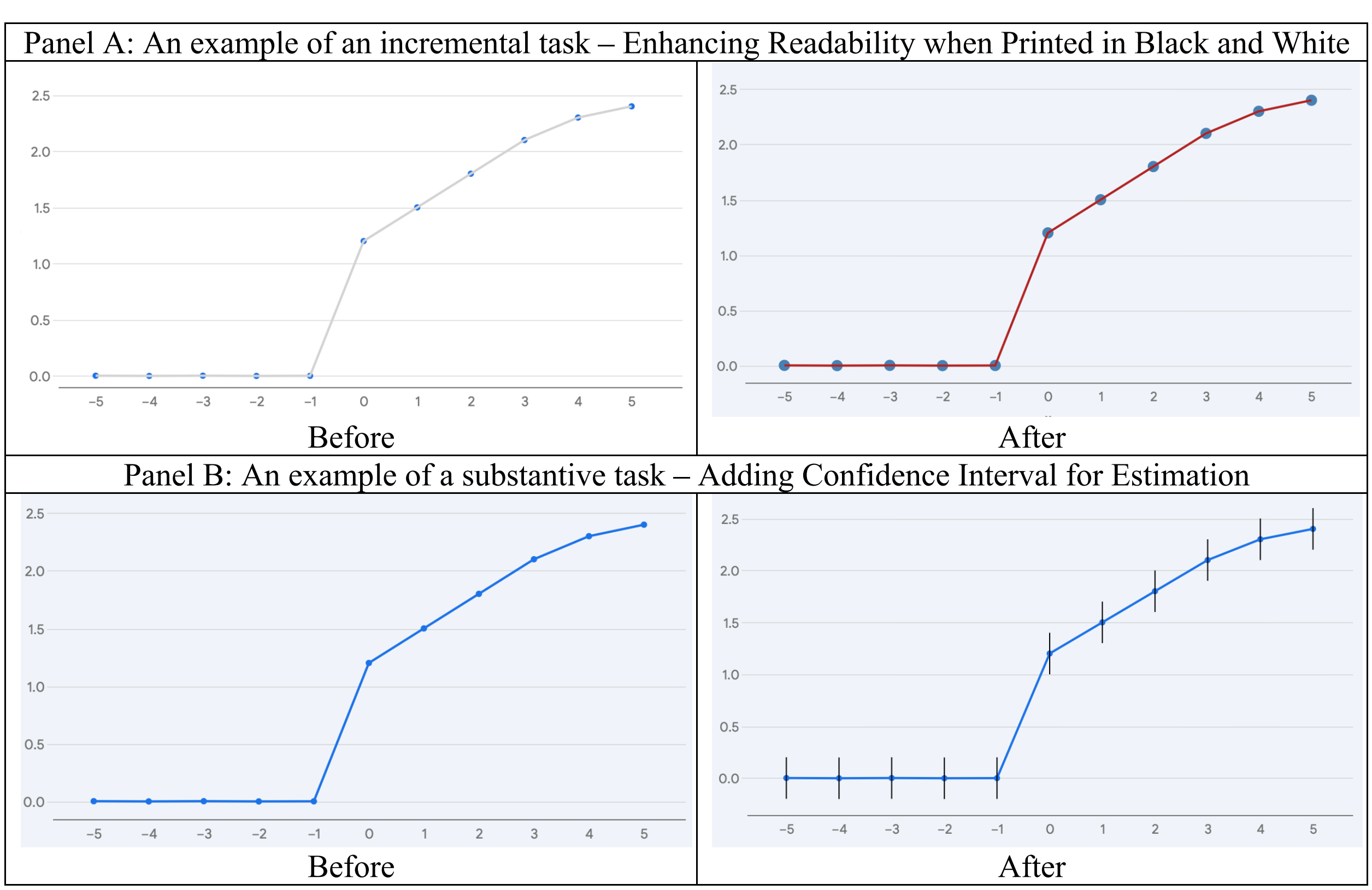

***Figure 1. Illustration of incremental and substantive innovation in a data visualization context.***

The cognitive-input distinction between substantive and incremental tasks generates a testable prediction. Because LLMs produce outputs by identifying patterns across large corpora of existing text and code (Brown et al. 2020), they should be more effective at tasks where the solution space is partly constrained by prior work. Incremental contributions fit this description: the problem is given, the existing codebase provides context, and the contributor's task is to identify the right local change. Substantive contributions fit it less well: the contributor must first identify what question to ask, and existing artifacts offer limited guidance for that decision. We therefore expect LLMs to disproportionately support incremental over substantive contributions. Testing this prediction in the field is difficult because voluntary adoption makes it hard to separate LLM effects from adopter characteristics. An ideal setting allows researchers to observe an LLM selectively supporting some collaborative projects but not all projects.

The launch of GitHub Copilot in October 2021 provides a rare natural experiment. GitHub Copilot, an LLM-powered coding assistance tool, initially supported certain languages, including Python and Rust, while omitting others, including R and Haskell. [2] These languages stand as major pillars of the open-source movement, with publicly accessible code repositories. The potential for innovation in these languages resides within their extensive ecosystems of user-contributed packages, such as pandas in Python or ggplot2 in R, that expand a language's capabilities and are maintained by communities of voluntary contributors. Because GitHub Copilot supported some languages like Python but not languages like R, we can compare the contribution trajectories of these communities with and without LLM assistance.

Most LLM studies examine closed-ended tasks in organizational settings and find benefits concentrated among lower-skilled workers (e.g., Brynjolfsson et al. 2025). Open-source contributors are highly skilled, participate voluntarily, and work without managerial direction. The few studies set in open-source environments (Chen et al. 2022; Hoffmann et al. 2024; Song et al. 2024; Liu et al. 2025) examine project/ developer-level outcomes without studying the composition type of individual contributions. Our first-order question (RQ1) asks, *do LLMs increase the volume of individual contributions in open-source settings*?

[2] OpenAI Codex, which powered GitHub Copilot, supported R and Haskell. However, the decision to exclude support was strategic, driven by the feasibility of corporate licensing and the viability of dedicated customer support.

Our primary contribution, however, concerns the composition of these contributions. Hence, we ask (RQ2), *how do LLMs affect the types of contributions in open-source innovation?* As we will show in what follows, open-source innovation needs both incremental contributions (e.g., code cleanup) and substantive contributions (e.g., feature development). Given the collaborative and self-selective nature of open-source projects, the differential effect of LLMs on these contribution types is ultimately an empirical question.

We also explore *how the effects of LLMs on contribution types may evolve as these models improve?* (RQ3). LLM effectiveness can grow through two channels: the context available to the model at inference time, and the model's own training. A useful way to study this is to examine how outcomes change when (a) the model is held constant, but the available context grows richer, and (b) model training itself improves. Together, these tests allow us to assess whether the gap between incremental and substantive contributions narrows, widens, or remains stable as LLMs get better.

We study these questions by comparing Python (supported by GitHub Copilot) with R (unsupported), two data-oriented languages with similar ecosystems focused on analytics, machine learning, and statistical computing. As a robustness check, we examine Rust (supported) and Haskell (unsupported), performance-oriented compiled languages emphasizing systems programming and functional paradigms.[3] Our main identification employs a Two-Way Fixed Effects (TWFE) model within a Difference-in-Differences framework using propensity score matching, with 1,187 Python packages matched to an equal number of R packages. We also estimate *Synthetic Difference-in-Differences* as an alternative specification. As a separate identification strategy that does not rely on cross-language comparisons, we employ a difference-in-differences approach *with temporal controls*, where the same Python packages serve as their own controls using data from an earlier period (Liaukonytė et al. 2023; Zhu et al. 2025). Our data covers all repository activity from October 2019 to December 2022, including over 1.1 million commits, with another 272 thousand Python commits between 2016-2018 for temporal control. The analysis period ends at the release of ChatGPT, which affects the control group. To classify contribution types, we undergo a time-intensive process of downloading all code-diffs to identify *function* additions, a proxy for substantive contributions. As an alternative measure, we use an LLM-based annotation method highlighted by Cheng et al. (2024) to classify commits into five categories aligned with established software engineering commit taxonomies (Hattori and Lanza 2008; Hindle et al. 2008): Maintenance, Code Development, Documentation and Style, Testing and Quality Assurance, and other non-programming tasks (details in Appendix C).

[3] We conduct combined analyses using all four languages together; results are reported in the robustness section.

To preview our results, we find a significant increase in the volume of voluntary contributions, with Python packages showing a 37% increase compared to matched R packages. The increase in incremental contributions (maintenance-related and feature-refining commits) is substantially larger than the increase in substantive contributions (code-development and function-introducing commits), and this difference is statistically significant. The in-context heterogeneity analysis shows that this disparity is more pronounced in projects where richer context is available to the LLM. The model-upgrade heterogeneity analysis shows significant effects over and above the main effects after GitHub Copilot's June 2022 upgrade, suggesting that as model training improves, the gap between incremental and substantive contributions widens further.

Our study makes several contributions. Identifying settings unaffected by generative AI is increasingly difficult, and our natural experiment around GitHub Copilot's selective language support offers one of the earliest field-level demonstrations that LLMs significantly boost voluntary contributions in collaborative environments. Second, we demonstrate that the cognitive-input structure of a task moderates how effectively LLMs can assist in collaborative settings. Incremental work draws on LLMs' capacity for contextual pattern-matching because existing code and prior contributions provide a reference the model can act on. Substantive work, which requires identifying what question to ask before a solution can be attempted, benefits less from this assistance. Finally, our analysis of context-rich projects and model upgrades suggests that as LLMs improve, the gap between incremental and substantive contributions widens rather than narrows. As incremental work becomes cognitively cheaper, the relative scarcity of substantive contributions grows, raising questions about how communities can incentivize contributors to invest in the deeper cognitive work that drives new directions in collaborative projects. This pattern carries implications for how collaborative innovation trajectories evolve across knowledge-intensive domains.

## 2. Literature Survey and Research Gap

### 2.1 Open-Source Innovation

Innovation carries different meanings across disciplines, yet nearly all definitions converge on the creation of new combinations of resources that people perceive as novel and improved (Rogers 2003). Open-source software development is organized as a private-collective innovation model (von Hippel and von Krogh 2003; Lin and Maruping 2022), one that puzzled economists because contributors invest significant effort without direct financial compensation (Lerner and Tirole 2002). Research has since identified two primary drivers: intrinsic satisfaction from solving problems and learning new skills, and extrinsic returns from skill signaling and reputation building within developer communities (von Krogh et al. 2012). Regardless of the motivation, individual investment generates collective value: participants pursue personal objectives while producing shared resources that any future contributor can use and build upon.

Several properties of open source make it a particularly informative setting for studying how new technologies affect innovation. Any contributor can understand and build upon the work of others without needing formal team membership (Lakhani and von Hippel 2004). Innovation occurs through a sequential layering of individual contributions, because modular architecture allows contributors to work on separable subtasks (Medappa and Srivastava 2019). Contributions are self-selected: contributors identify opportunities and gravitate toward tasks that align with their motivations and expertise (Miric and Jeppesen 2023; Piezunka and Dahlander 2019). Contributors also use open-source work to sharpen skills that they cannot practice in organizational settings, making this self-selection also a learning decision (Shah 2006).

These properties together create an environment where LLMs could play out differently than in organizational settings. If LLMs make certain types of contributions cognitively cheaper, contributors may shift toward those tasks, away from them, or use the freed-up effort to attempt harder problems. How these properties interact with LLM capabilities to reshape the mix of contributions is an open question.

### 2.2 Substantive and Incremental Innovation

Innovation tasks in knowledge contexts can be conceptualized along a continuum, though scholarly frameworks have repeatedly anchored them around dichotomous constructs: radical versus incremental innovation (Tushman and Anderson 1986), architectural versus component innovation (Henderson and Clark 1990), exploration versus exploitation (March 1991), and degrees of departure from existing knowledge (Gatignon et al. 2002). These typologies share a common focus on product innovation in organizational settings. They say less about collaborative processes in which contributors self-select into tasks based on cognitive demands rather than managerial assignment.

Substantive innovation focuses on expanding what a project can accomplish by adding new functionality (Huang et al. 2026), requiring "question ingenuity": the cognitive capacity to identify what question is worth asking and then to solve it. This conception draws on exploration processes (March 1991; Benner and Tushman 2003). The problem space is less specified, the solution is not predetermined, and the contributor must invest considerable effort to define both. In software development, substantive innovation typically manifests as implementing new functionalities that extend what a project can accomplish.

Incremental innovation centers on improving the quality of existing elements, making them function better, more reliable, or more efficient. The primary cognitive demand is comprehension of the current context before making targeted changes, drawing on exploitation activities (March 1991; Benner and Tushman 2003; Medappa and Srivastava 2019). These tasks operate within a more clearly defined problem and solution space. In software contexts, incremental innovation includes bug fixes, code cleanup, and refactoring, activities that Hattori and Lanza (2008) classify as maintenance-oriented.

It is important to note that the axis separating substantive from incremental innovation is the structure of cognitive input required, not the importance of the outcome. An incremental contribution, such as fixing a critical security bug, can be more valuable than adding a rarely used feature. What differs is how specified the problem space is at the outset and how much cognitive effort goes into defining the question versus executing within known parameters. Table 1 positions this distinction alongside existing typologies.

The emergence of LLMs creates a natural question about their differential impact on these two types (Lin and Maruping 2025). Because LLMs are effective at in-context learning, generating solutions from existing patterns and examples (Brown et al. 2020), they may disproportionately reduce the cognitive cost of incremental contributions where the problem space is already defined. Whether they similarly support substantive contributions, where the problem itself must be formulated, is less clear. We develop this reasoning into specific research questions below.

**Table 1. Positioning Substantive/Incremental Innovation among Extant Theoretical Constructs**

| Theoretical Construct | Focus and Context | Seminal Work | Context and Emphasis |
|---|---|---|---|
| Radical/Incremental Innovation | Focuses on degree of technological change and disruption to existing competencies | Tushman and Anderson (1986) | Primarily examines organizational and market impacts rather than cognitive processes and human input requirements. |
| Competence Acquisition/ Enhancement | Examines changes to firm capabilities | Gatignon et al. (2002) | Focuses on organizational outcomes, not on cognitive processes or human input requirements |
| Exploration/ Exploitation | Focuses on organizational learning and knowledge balance | March (1991); Benner and Tushman (2003) | Broader scope beyond innovation, including organizational adaptation. |
| Architectural/ Component Innovation | Distinguishes between changes to system linkages vs. components | Henderson and Clark (1990) | Focuses on technical structure rather than cognitive processes. |

### 2.3 Research Questions

Given that external events such as technology acquisitions can alter contribution patterns in open-source projects (Chen et al. 2022), our first question studies whether LLMs change the total volume of open-source contributions. Extant work documents productivity gains from LLMs in controlled organizational settings (e.g., Brynjolfsson et al. 2025), but the voluntary nature of open source complicates prediction. If LLMs reduce effort per contribution, contributors may produce more. Alternatively, if automation diminishes the learning and signaling returns that motivate voluntary participation (Lerner and Tirole 2002), contributors may reduce effort or disengage. The direction is not obvious ex ante:

*RQ1: How do LLMs affect the volume of contributions in open-source innovation?*

Our second and primary question concerns the composition of contributions. While emerging work examines overall productivity in open-source contexts (Song et al. 2024; Hoffmann et al. 2024), the cognitive-input framework in Section 2.2 points to a more specific question about composition. Incremental contributions operate within a defined problem space: the contributor must comprehend existing code and make targeted changes. This is the type of task where in-context learning (Brown et al. 2020) should be most useful, because the existing codebase provides the patterns and the solution space is bounded. Substantive contributions require formulating the problem itself, a cognitive task where LLM assistance is less direct. If LLMs disproportionately reduce the cost of incremental work, the composition of innovation may shift toward it. Yet this reasoning also admits an alternative: LLMs may free contributors from routine tasks, allowing them to reallocate effort toward substantive work. The empirical direction remains open:

> *RQ2: How do LLMs differentially affect substantive versus incremental contributions in open-source innovation?*

Our third question explores whether the relationship between LLMs and contribution types evolves. Two channels could drive such evolution. First, the availability of project context at inference time: in projects with more extensive codebases and documentation, LLMs can draw on richer patterns when generating suggestions. If incremental tasks benefit more from contextual richness, the differential effect may be amplified in context-rich settings. Second, improvements in model training: as LLMs are trained on larger and more diverse corpora, their capacity to support less-specified tasks may grow, potentially narrowing the gap between their impact on incremental and substantive work:

> *RQ3: How does the differential impact of LLMs on contribution types vary with context availability and model capability?*

## 3. Research Context and Data

### 3.1 Empirical Context: The Natural Experiment

The natural experiment that our investigation exploits is the launch of GitHub Copilot, an LLM-powered coding assistance tool. Developed by GitHub and OpenAI, Copilot was launched in October 2021 with selective language support: Python was supported from the start, while R was not. At least until the ChatGPT launch in late 2022, R packages did not have access to mainstream LLM coding assistance (see Figure 2). This selective support creates a natural experiment, allowing us to compare contribution trajectories in Python (treated) and R (control) packages before and after Copilot's introduction.

Our primary comparison contrasts community contributions to open-source packages across these two languages. Both Python and R are primarily data-oriented languages with large, active developer communities. Each ecosystem is supported by a wealth of libraries and frameworks (i.e., packages) such as Python's NumPy and pandas, and R's ggplot2 and dplyr. The open-source communities for these languages are hosted on GitHub, a platform central to the ecosystem of open-source projects with over 100 million developers contributing across more than 500 programming languages. On GitHub, communities organize around code "repositories," where contributors propose and implement code changes through "commits." Each commit is a record of a modification to the source code, capturing the specific changes made, along with metadata such as the author's name, a timestamp, and a message describing the purpose of the change. As a robustness check, we extend the analysis to a second language pair, Rust (treated) versus Haskell (control), two performance-oriented languages that share a focus on type safety and systems programming, providing an independent replication outside the data-science domain (Section 5.4.1).

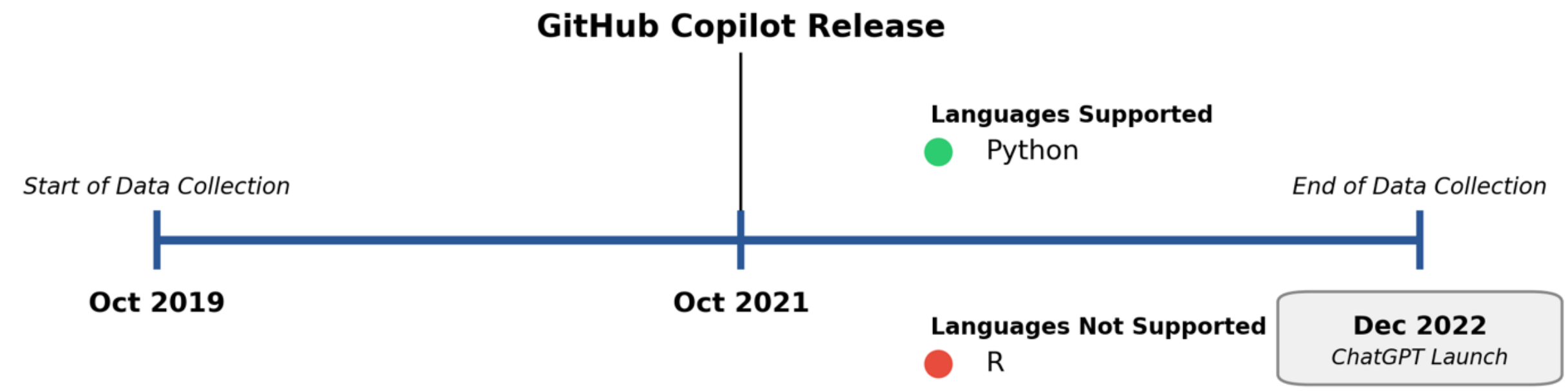

***Figure 2. Timeline of the natural experiment***

## 3.2 Data Collection

We collected data from October 2019 to October 2022, a period spanning two years before and one year after the introduction of GitHub Copilot. For Python packages, we focused on the 2,000 most popular packages as of 2021.[4] We obtained the entire list of R packages from R's central repository, the Comprehensive R Archive Network (CRAN), as we did not find a reliable index of well-maintained packages. We excluded 348 Python packages and 7,965 R packages that had not been updated in the two years preceding Copilot's release, as they were unlikely to be influenced by it. After matching (described

[4] van Kemenade et al. (2021) aggregate and rank packages monthly on the top-pypi-packages repository. Crates.io/crates maintain the ranked list for Rust, and hackage.haskell.org maintains the ranked list for Haskell.

in Section 4.1), our primary dataset comprises 1,187 matched Python/R package pairs, aggregated at a quarterly level to account for the cyclic nature of package releases. For the robustness analysis, we follow the same procedure for Rust and Haskell, starting from the top 2,000 Rust packages on *Crates.io* and the top 2,000 Haskell packages on *Hackage*, yielding 1,373 matched pairs. Table 2 provides variable descriptions, and Table A1 provides summary statistics.

**The Commits Dataset**: We used GitHub's official API to programmatically download commits from the repositories associated with each package. When developers "commit" a code change, they save the current state of their project, accompanied by a message describing the changes. Our primary Python/R dataset encompasses over 620,000 commits. An additional 273,000 Python commits from 2016 to 2018 serve as temporal controls (described in Section 4), and approximately 314,000 Rust and Haskell commit.

**Table 2. Variable Explanation**

| Variable | Description |
| --- | --- |
| $Commits_{it}$ | The number of new commits of the package *i* in the quarter *t* |
| $Commits_{ijt}^{v}$ | The number of new commits of type *j* of the package *i* in the quarter *t*,<br>when *v*=LLM, j ∈ (Maintenance, Code Development)<br>when *v*=Function, j ∈ (With function definition, Without function Definition). |
| $VersionRelease_{it}$ | The number of version releases of the package *i* in the quarter *t* |
| ***DiD* Variables** | |
| $TreatmentLang_i$ | A binary variable that equals 1 if package *i* is from Python (or Rust). |
| $afterCopilot_t$ | A binary variable that equals 1 if quarter *t* occurs after October 2021, i.e., after GitHub Copilot's launch |
| **Triple Difference Variables** | |
| $ImprovedModel_t$ | A binary variable that equals 1 for quarters after June 2022 (GitHub Copilot's significant model improvement |
| $ProjectActivity_i$ | A binary variable that equals 1 if package *i* had above-median activity |

**The Version Release Dataset:** For each package, we collected version release dates from official repositories: PyPI for Python and CRAN for R (and Crates.io for Rust and Hackage for Haskell).

### 3.3 Classifying Contribution Types

The theoretical distinction between substantive and incremental innovation (Section 2.2) is operationalized at the commit level using two complementary approaches. The first is a code-based approach, which examines the code itself: it tags a commit as substantive if it introduces a new function, and as incremental otherwise, on the logic that new features are typically delivered as new functions while edits to existing code refine existing functionality. The second is an LLM-based approach, which examines the commit message: using annotation guidelines described in Section 3.3.2, an LLM assigns each commit to one of five categories, two of which map directly to substantive and incremental contributions. The two approaches draw on different signals (code structure vs. developer-written description) and serve as independent checks on the same theoretical construct.

### 3.3.1 Code-Based Approach: Function Definition Detection

Our first approach identifies substantive contributions by examining *code diffs* [5] to determine when new functions are introduced. We analyze all code changes in each commit by processing the difference between previous and current versions of files modified in that commit.[6] We then count the number of functions added minus functions removed across all files in a commit. When an existing function is edited, it appears in code diffs as one function removed and one function added. To account for this, we classify a commit as a substantive contribution only if there is a positive net addition of functions across all files in the commit (With Function Definition => 1). Otherwise, the commit is classified as an incremental contribution. Function detection is implemented with language-specific parsers; full details are in Appendix D.

### 3.3.2 LLM-Based Approach: Commit Message Classification

Our second approach identifies substantive contributions by examining the description in the commit message written by the author. Classifying over one million GitHub commits into specific categories is a challenging task, given the absence of pre-defined criteria and established categorization for commits. GitHub commit messages are often unstructured and do not explicitly convey the type of change introduced. Understanding the types of contributions by developers is, however, central to understanding whether and how LLMs affect the trajectory of open-source projects. To address this challenge, we employed an LLM-based annotation method, as recommended by Cheng et al. (2024), for the classification of commits.

---

[5] A "code diff" is the line-by-line record of additions and deletions that a version control system generates when a commit modifies a file.

[6] Downloading the code diff as well as the actual edited file is an extremely space and time-intensive process, taking over 4 months and close to 0.5 TB of code download using GitHub Official API.

***Category Identification and Refinement:*** We randomly selected 500 Python and 500 R commit messages and asked GPT-4 to assign categories without providing explicit criteria, asking the model to justify each choice. This resulted in approximately 250 categories, many of which overlapped. To consolidate these into broader groupings, we relied on both GPT-4's language capabilities and independent judgment from a human expert research assistant. The RA aggregated categories based on the LLM's reasoning, their own interpretation of the commit messages, and established commit classification taxonomies (Hattori and Lanza 2008; Hindle et al. 2008). This process yielded five mutually exclusive categories: "*Maintenance*" (bug fixes, code cleanup, library dependency updates, deprecation, refactoring), "*Code Development*" (feature addition, code optimization), "*Documentation and Style*" (documentation, style updates), "*Testing and Quality Assurance*" (testing, automatic version updates), and "*Other*" (license and copyright updates).[7]

***Annotation Guidelines and Validation:*** We developed detailed annotation guidelines to categorize commits into five categories. Appendix Table C1 presents the annotation guidelines used with the LLMs.

***LLM Benchmarking and Selection:*** We randomly selected 200 Python and 200 R commit messages and had them tagged by three expert human annotators. Using this human baseline, we benchmarked five models (GPT-4o, GPT-4o mini, LLaMA 3.3 70B Q8.0, DeepSeek R1, and DeepSeek V3) against human annotations. Figure 3 shows the agreement levels between expert annotators and each LLM.

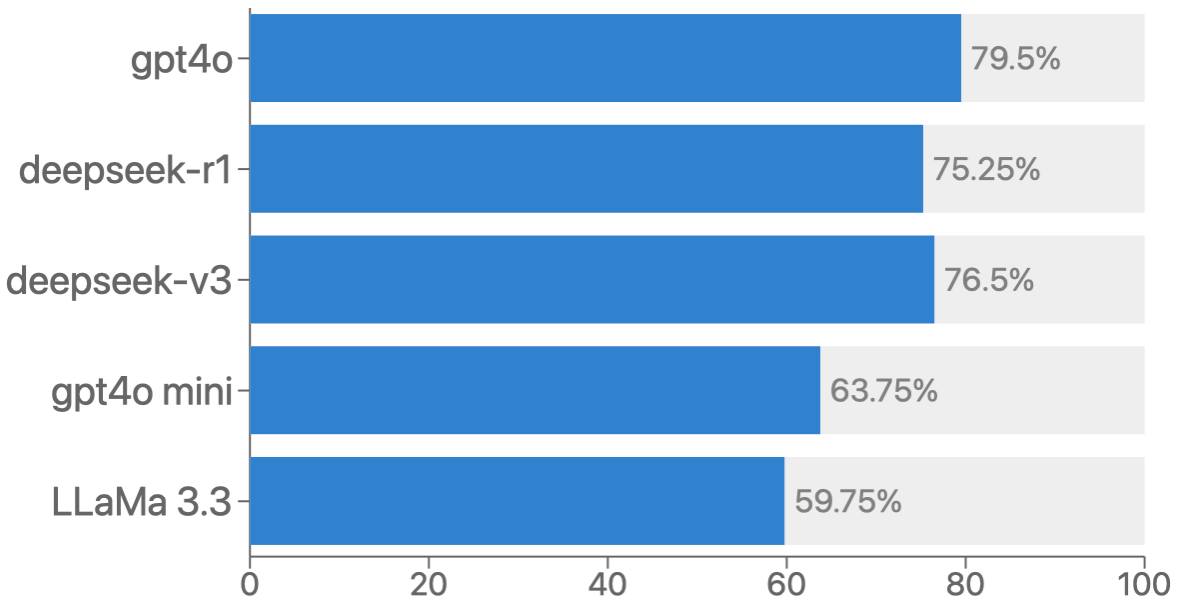

**Figure 3.** The accuracy of LLMs compared to expert human annotation.

Cohen's Kappa exceeded 0.6 for GPT-4o and DeepSeek, placing agreement at the "substantial" level (Cohen 1960; McHugh 2012). Although DeepSeek was cheaper, its throughput (in 2024) of approximately

[7] For instance, "this commit fixes a bug in the error handling module" got tagged as Maintenance, "added a new feature for data visualization" as Code Development, "updating the README file with usage instructions" as Documentation and Style, "unit tests for the fixes introduced in #c1287" as Testing and Quality Assurance, and "updating the license file to the latest version" as Other.

80 tokens per second made it impractical for processing the roughly 720 million tokens in our corpus. We employed GPT-4o.

# 4. Identification Strategies

We employ three complementary and independent identification strategies in our investigation.

## 4.1 Classical Propensity Score Matching with Difference in Differences Technique

Given that our natural experiment has a well-defined treated group (Python) and control group (R) with one-shot policy adoption, we employ the classic Difference-in-Differences (DiD) framework (Meyer 1995) and estimate the Average Treatment Effect on Treated (ATT) using a two-way fixed effects (TWFE) model. One potential threat to causal estimation using DiD in a natural experiment setting is that the assignment of treatment could be non-random. In our context, while the assignment of treatment (i.e., language supported by GitHub Copilot) was a business decision and is therefore exogenous to the package themselves, the choice for the package to be on a programming language is likely endogenous. To address the potential selection, we employed the *Propensity Score Matching* to pre-process the two package groups (Dehejia and Wahba 2002). Specifically, we use the nearest neighbor method without replacement to obtain one control package for every treated package (e.g., Mayya and Viswanathan 2025). We matched based on their pre-treatment commit activities, to ensure the robustness of matching based on different types of pre-treatment activities (i.e., package release and commit activities). Our primary dataset comprises 1,187 matched Python/R package pairs. The same procedure yields 1,373 matched Rust/Haskell pairs for the robustness analysis (Section 5.4.1). We formally set up the model as follows:

$$Y_{it} = \alpha_i + \gamma_t + \beta\ TreatmentLang_i\ X\ afterCopilot_t + \eta_{it} \qquad (1)$$

where $Y_{it}$ is the dependent variable of interest (the count of commits) for package *i* in quarter *t*. In the model, $\alpha_i$ is the package-specific fixed effects; $\gamma_t$ is the time-period fixed effects; $TreatmentLang_i$ carries the value of 1 if the package $i$ belongs to the language supported by GitHub Copilot at launch (Python) and $AfterCopilot_t$ denotes the time after the launch of GitHub Copilot (t ≥ 0) We cluster the standard errors by package in the TWFE analyses. The coefficient of interest is $\beta$, which estimates the difference between the treatment and control packages after the launch of Copilot.

### 4.1.1 Testing the Parallel Trends

To ensure that the treated and control packages have a similar trend before the release of GitHub Copilot, we employ the lead-lag model proposed by Autor (2003). Specifically, we treated the quarter before the launch of GitHub Copilot as the baseline and estimated the lead-lag model as follows:

$$Y_{it} = \alpha_i + \gamma_t + \sum \alpha_j^p \ TreatmentLang_i \ X \ RelativeQuarter_j + \eta_{it} \qquad (2)$$

where, *RelativeQuarter*$_j$ is a vector of dummies that indicates Quarter $j$ before and after the treatment took place. As per Autor (2003), the parallel trends assumption holds if all the coefficients ($\alpha_j^p$) are not statistically significant for relative quarters $j$ prior to the treatment. Figure 4 visualizes the graph and demonstrates that the parallel trends assumption is not violated (Appendix Figure B1 for Rust/Haskell).

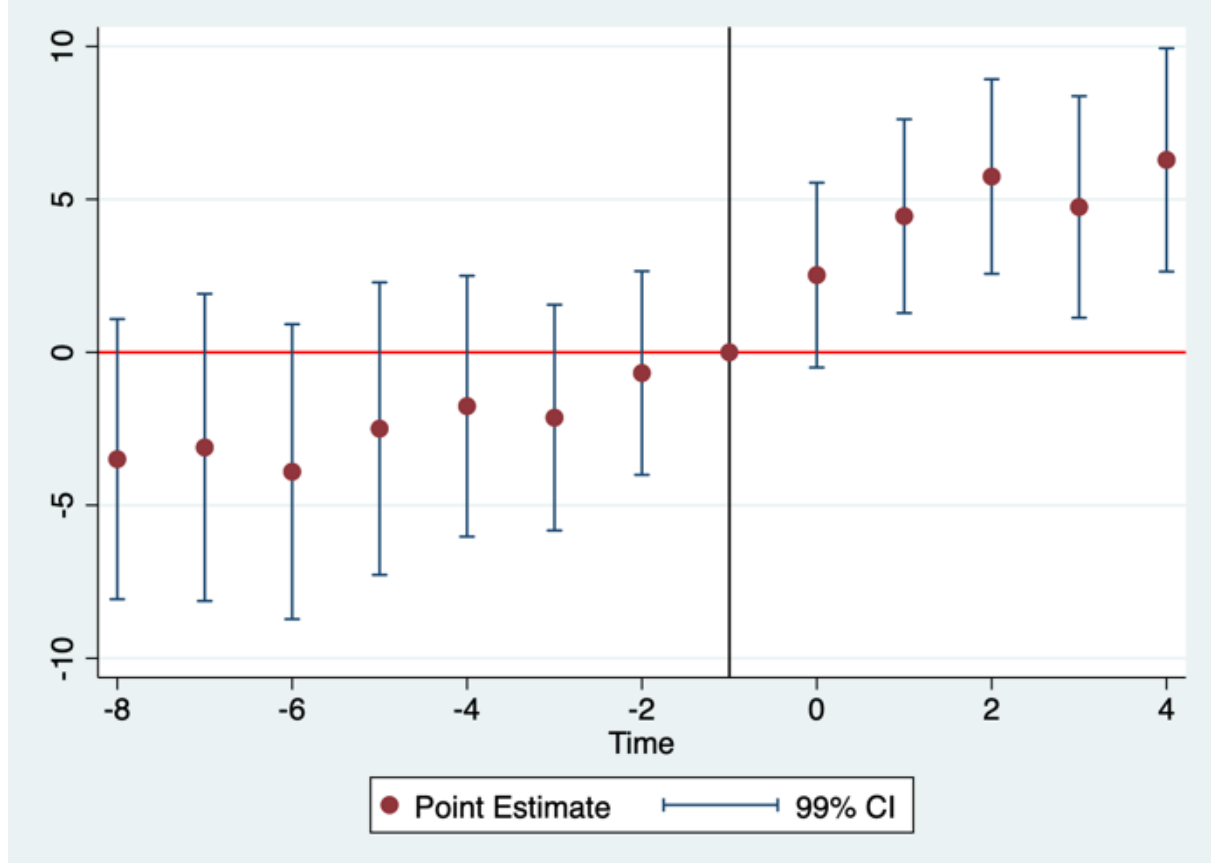

Python vs. R Parallel Trends
**Figure 4**. Parallel Trends Graph.
Note: The y-axis is the raw value of the commit count each quarter. 99% confidence interval bars.

## 4.2 Synthetic Difference in Differences (SDiD) as an Alternative Identification Strategy

We extend our empirical analysis by employing Synthetic Difference-in-Differences (SDiD), a recent estimation tool introduced by Arkhangelsky et al. (2021). SDiD builds upon the Synthetic Control technique (Abadie et al. 2010) by creating a "synthetic" control group. SDiD also generates weights to time-periods such that observations closer to the treatment periods receive higher weights compared to observations which are a few quarters away. This has been successfully applied in several recent studies (Berman and Israeli 2022; Lambrecht et al. 2023). Formally, we follow Arkhangelsky et al. (2021) and analyze a balanced panel with $N$ units and $T$ time periods, where the outcome for unit *i* in period *t* is denoted by $Y_{it}$, and exposure to the binary treatment, the release of GitHub Copilot, is denoted by $W_{it} \in \{0,1\}$. The method first finds unit weights $\omega_i$ that align pre-exposure trends in the outcomes of unexposed units with those of the exposed units. The model also generates time weights $\lambda_t$ that balance the pre-exposure time periods with post-exposure ones, with observations closer to the treatment period receiving higher weights than observations further away. Formally, the SDiD estimation procedure solves:

$$(\hat{\tau}, \hat{\mu}, \hat{\alpha}, \hat{\beta}) = argmin\,\{\sum_{i=1}^{N} \sum_{T=1}^{t} (Y_{it} - \mu - \alpha_i - \beta_t - TreatmentLang_i \mathrm{X} AfterCopilot_t \tau)^2 \widehat{\omega_\iota, \lambda_t}\} \quad (3)$$

where $\tau$ is the ATT. The dependent variable, $Y_{it}$ , is the variable of interest (such as the number of new versions) for package *i* in a specific quarter relative to the treatment period, *t*. Here, the model employs both the package-fixed effects and the period-fixed effects and $TreatmentLang_i \mathrm{X} AfterCopilot_t$ denotes treatment language package *i* in the periods after the launch of GitHub Copilot. Since the adoption is not staggered, equation (3) estimates a TWFE DiD model identical to Equation (1) with the addition of unit-specific weights $\omega_i$ and time-specific weights $\lambda_t$. The unit weights $\omega_i$ are selected such that the weighted pre-treatment control outcomes have a similar trend to that of the average outcomes of the treatment units:

$$\widehat{\omega_0} + \sum_{i=1}^{N_{co}} \widehat{\omega_\iota}\, log\, log\, (y_{it} + 1) \approx \frac{\sum_{i=N_{co}+1}^{N} log\, log\, (y_{it} + 1)}{|N_{tr}|}$$

To estimate the standard errors, we use the Jackknife Variance Estimation. The time weights $\lambda_t$ are constructed by making the pre-treatment time periods similar to the treatment periods for the control group:

$$\widehat{\lambda_0} + \sum_{t=-8}^{t=-1} \widehat{\lambda_t}\, log\, log\, (y_{it} + 1) \approx \frac{\sum_{i=N_{co}+1}^{N} log\, log\, (y_{it} + 1)}{T_{post}}$$

### 4.3 Difference-in-Differences with Temporal Controls

Our primary identification relies on cross-language comparisons. To address any concern about unobserved differences between Python and R ecosystems, we employ a complementary strategy that does not rely on cross-language comparisons: a difference-in-differences approach with temporal controls, where the same

Python packages serve as their own controls using data from an earlier time period (Liaukonytė et al. 2023; Zhu et al. 2025). Specifically, we compare the same Python packages across two windows: a control period (2016 Q2 to 2018 Q4) and a treatment period (2020 Q2 to 2022 Q4), with perfect calendar-quarter alignment[8]. The design incorporates era-specific package fixed effects ($\alpha_i^{control}$ and $\alpha_i^{treated}$) to allow for secular changes in baseline commit activity, and relative-time fixed effects ($\gamma_t$) to control for seasonal patterns. This temporal control approach complements our primary cross-language specification by exploiting within-package variation over time rather than across programming languages. The 273,000 Python commits from the control period were classified using the same approaches described in Section 3.3. Figure 5 shows that the parallel trends for this specification hold across all contribution types.

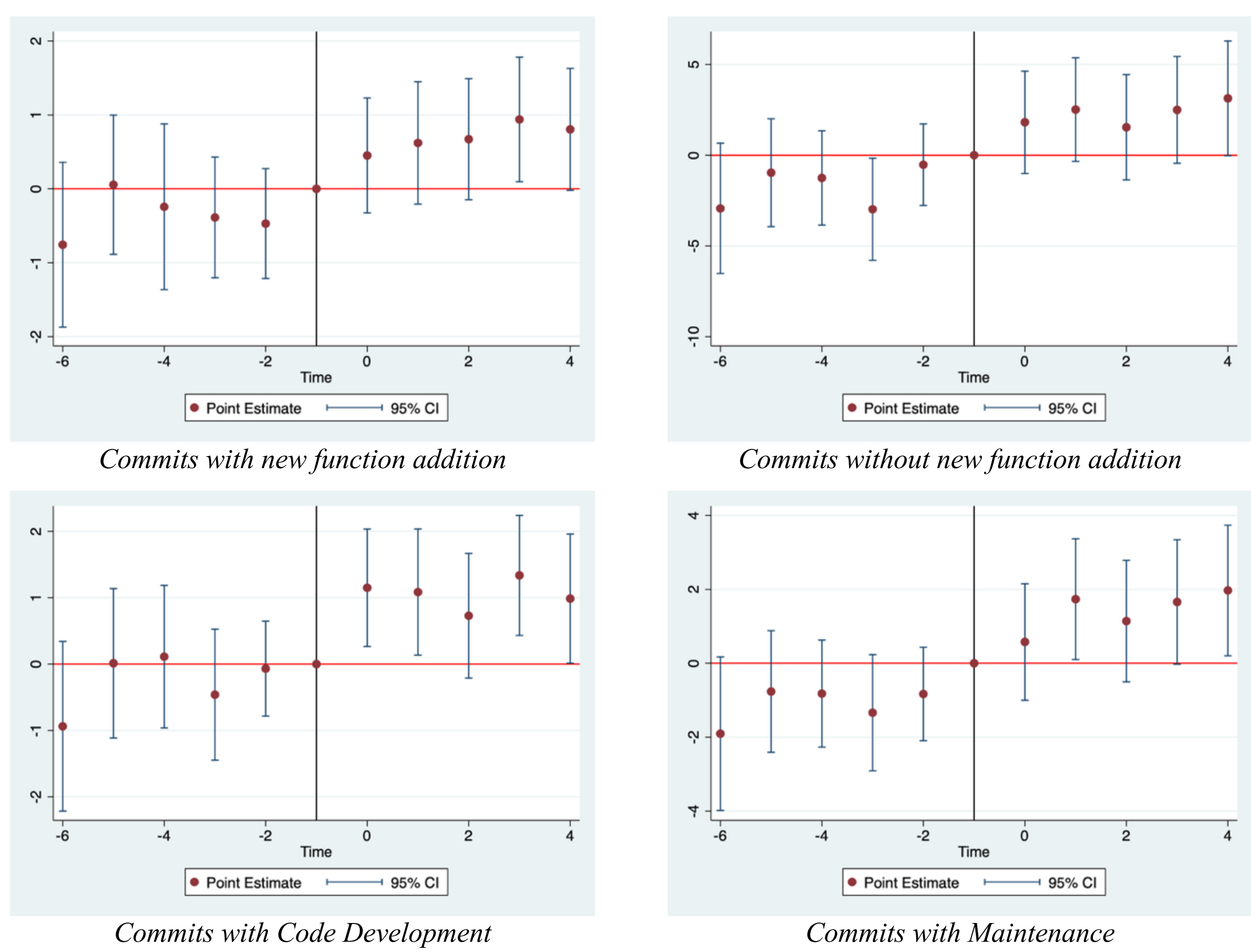

*Commits with new function addition*

*Commits without new function addition*

*Commits with Code Development*

*Commits with Maintenance*

**Figure 5.** Parallel Trends for Difference-in-Differences with Temporal Controls.

[8] Q4 2015 and Q1 2016 were extreme high activity time for Python communities for primarily two reasons. First, Python 3.5, the first production-ready Python 3 version with certain features (like await, async, matrix multiplication) only on Python 3, pushed all communities to provide support for Python 3. Second, Google open-sourced TensorFlow, which resulted in massive downstream adoption and support, especially from the data-science community.

## 5. Results

### 5.1 *The Effect of GitHub Copilot on the Volume of Contributions*

The first research question asks how GitHub Copilot affects the volume of open-source contributions, measured as the count of commits per package per quarter. Table 3 presents results estimated using three specifications: TWFE (Columns 1 and 4), balanced panel TWFE (Columns 2 and 5), and Synthetic DiD (Columns 3 and 6). The cross-language comparison of Python versus R (Columns 1-3) shows that GitHub Copilot increases commits by 6.816 per quarter in the TWFE specification ($p<0.001$), a 39.55% increase relative to the pre-treatment mean of the treated group of 17.235 commits per quarter (Angrist and Pischke 2009). The effect is consistent across specifications: 5.496 commits (31.89%) in the balanced panel and 4.894 commits (28.40%) in synthetic DiD. The temporal control design (Columns 4-6), which compares the same Python packages across a pre-Copilot era and a treatment era (Section 4.3), produces convergent evidence: 4.638 additional commits per quarter ($p<0.01$) in TWFE, a 27.13% increase relative to the treated group's pre-treatment mean of 17.097 commits. The balanced panel shows 4.328 commits (25.31%, $p<0.05$) and synthetic DiD shows 3.431 commits (20.07%, $p<0.05$). The agreement between these two identification strategies, one exploiting cross-language variation and the other within-language temporal variation, strengthens the causal interpretation of the volume effect.

**Table 3. The Effect on the Volume of Innovation**
**Number of New Commits**

| Comparison | Python (Treated) vs. R (Control) | | | Temporal Controls (Python) | | |
|---|---|---|---|---|---|---|
| Models | (1) *TWFE* | (2) *Balanced TWFE* | (3) *SynthDiD* | (4) *TWFE* | (5) *Balanced TWFE* | (6) *SynthDiD* |
| *TreatmentLang*$_i$ *X afterCopilot*$_t$ | 6.816*** (0.892) | 5.496*** (0.941) | 4.894*** (0.907) | 4.638** (1.624) | 4.328* (1.874) | 3.431* (1.732) |
| Time Fixed Effect | YES | YES | YES | YES | YES | YES |
| Package Fixed Effect | YES | YES | YES | YES | YES | YES |
| # of Obs. | 29,050 | 24,739 | 24,739 | 22,677 | 19,426 | 19,426 |

Note: Standard errors clustered around package in parentheses for TWFE models. The variance for SynthDiD is estimated using the jackknife variance procedure. Columns 1-3 compare 1,187 matched Python and R packages. Columns 4-6 compare the same Python packages across a control period (2016 Q2 to 2018 Q4) and a treatment period (2020 Q2 to 2022 Q4), as described in Section 4.3.
*** $p<0.001$, ** $p<0.01$, * $p<0.05$

Figure 6 visualizes the synthetic DiD analyses. Panel (a) shows the cross-language comparison of Python versus R. The turquoise line represents the trend for Python packages in the post-treatment period, the shaded red area delineates the synthetic control group constructed from R packages, and the dashed black line indicates the counterfactual for Python had GitHub Copilot not been introduced. Panel (b) shows the temporal control design, comparing the same Python packages across a pre-Copilot and a treatment era.

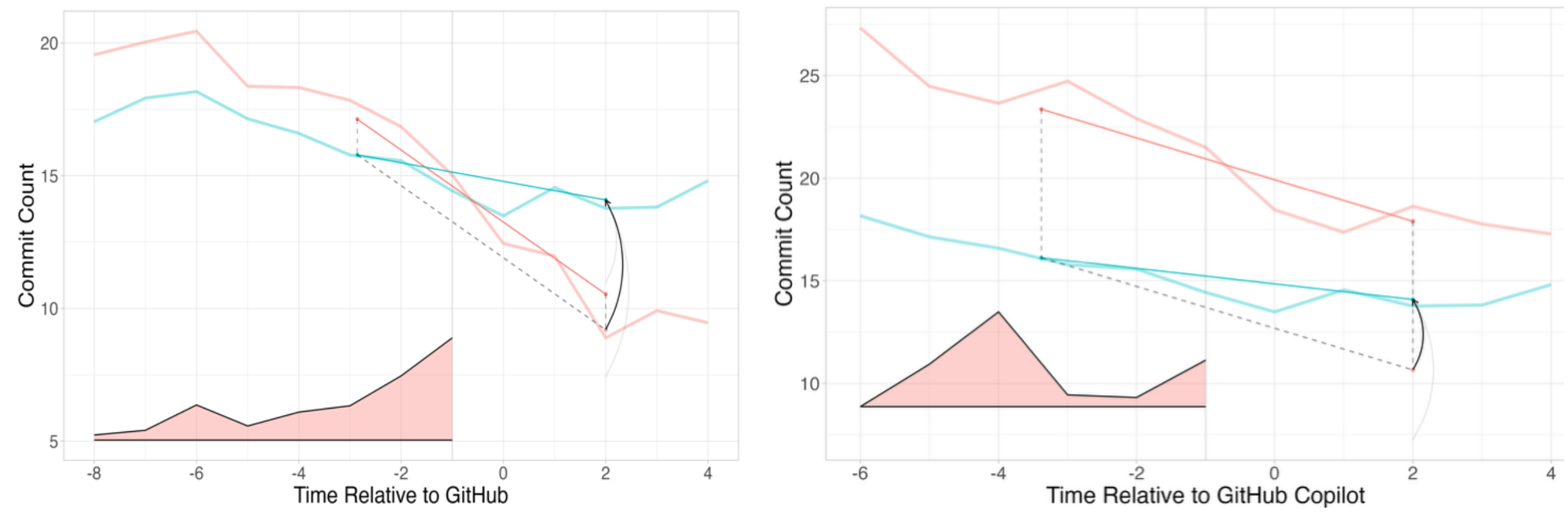

**Figure 6a. Python vs. R.** **Figure 6b. Temporal Controls (Python).**

Note: The x-axis indicates the relative period to the launch of GitHub Copilot (October 2021)

## 5.2 The Effect of GitHub Copilot on the Type of Contributions

Having established that Copilot increases the total volume of contributions, we now examine whether this increase differs across contribution types. We operationalize the distinction between substantive contributions (higher cognitive input) and incremental contributions (lower cognitive input) through two approaches: structural analysis of function additions in code diffs (Section 5.2.1) and semantic classification of commit messages using an LLM (Section 5.2.2). For both approaches, we estimate equation (1) separately for each contribution type, replacing the DV with $\text{Commits}_{ijt}^{v}$, where v represents the categorization approach, i indexes the package, t indexes the quarter, and j indexes the contribution type.

### 5.2.1 Function Analysis

Panel A of Table 4 reports results from the cross-language comparison of Python versus R. For commits that add new functions (Columns 1-3), our TWFE estimation shows an increase of 0.800 commits per package-quarter, with consistent effects in the balanced panel (0.626) and synthetic DiD (0.600), all significant at $p < 0.001$. For commits without new function additions (Columns 4-6), we observe larger effects: 6.016 additional commits per quarter in TWFE, 4.870 in the balanced panel, and 4.226 in synthetic DiD, again all significant at $p < 0.001$. The difference between the two contribution types is statistically significant in triple-difference specifications (Appendix Table A2). Appendix Figure A1 visualizes the Synthetic DiD estimation. Panel B of Table 4 replicates this analysis using our temporal controls design. The results are directionally consistent: both contribution types increase, with larger effects for commits without new functions. Figure 7 visualizes the coefficient estimates.

**Table 4. The Effect on the Type of Contributions: Function Addition**

**Panel A: Python (Treatment) vs. R (Control) Analysis**

| | New Functions Added | | | New Functions Not Added | | |
|---|---|---|---|---|---|---|
| Dependent Variable | (1) *TWFE* | (2) *Balanced TWFE* | (3) *SynthDiD* | (4) *TWFE* | (5) *Balanced TWFE* | (6) *SynthDiD* |
| $TreatmentLang_i$ X $afterCopilot_{it}$ | 0.800*** (0.190) | 0.626*** (0.188) | 0.600*** (0.172) | 6.016*** (0.761) | 4.870*** (0.806) | 4.226*** (0.787) |
| Time Fixed Effect | YES | YES | YES | YES | YES | YES |
| Package Fixed Effect | YES | YES | YES | YES | YES | YES |
| # of Obs. | 29,050 | 24,739 | 24,739 | 29,050 | 24,739 | 24,739 |

**Panel B: Temporal Controls (Python)**

| | New Functions Added | | | New Functions Not Added | | |
|---|---|---|---|---|---|---|
| Dependent Variable | (1) *TWFE* | (2) *Balanced TWFE* | (3) *SynthDiD* | (4) *TWFE* | (5) *Balanced TWFE* | (6) *SynthDiD* |
| $TreatmentLang_i$ X $afterCopilot_{it}$ | 1.069* (0.439) | 0.795 (0.490) | 0.575+ (0.341) | 3.606** (1.323) | 3.511* (1.534) | 2.722+ (1.400) |
| Time Fixed Effect | YES | YES | YES | YES | YES | YES |
| Package Fixed Effect | YES | YES | YES | YES | YES | YES |
| # of Obs. | 22,677 | 19,426 | 19,426 | 22,677 | 19,426 | 19,426 |

Note: Standard Errors clustered around package in parentheses for columns (1), (2), (4) and (5). The variance for (3) and (6) are estimated using the Jackknife variance procedure.
*** p<0.001, ** p<0.01, * p<0.05

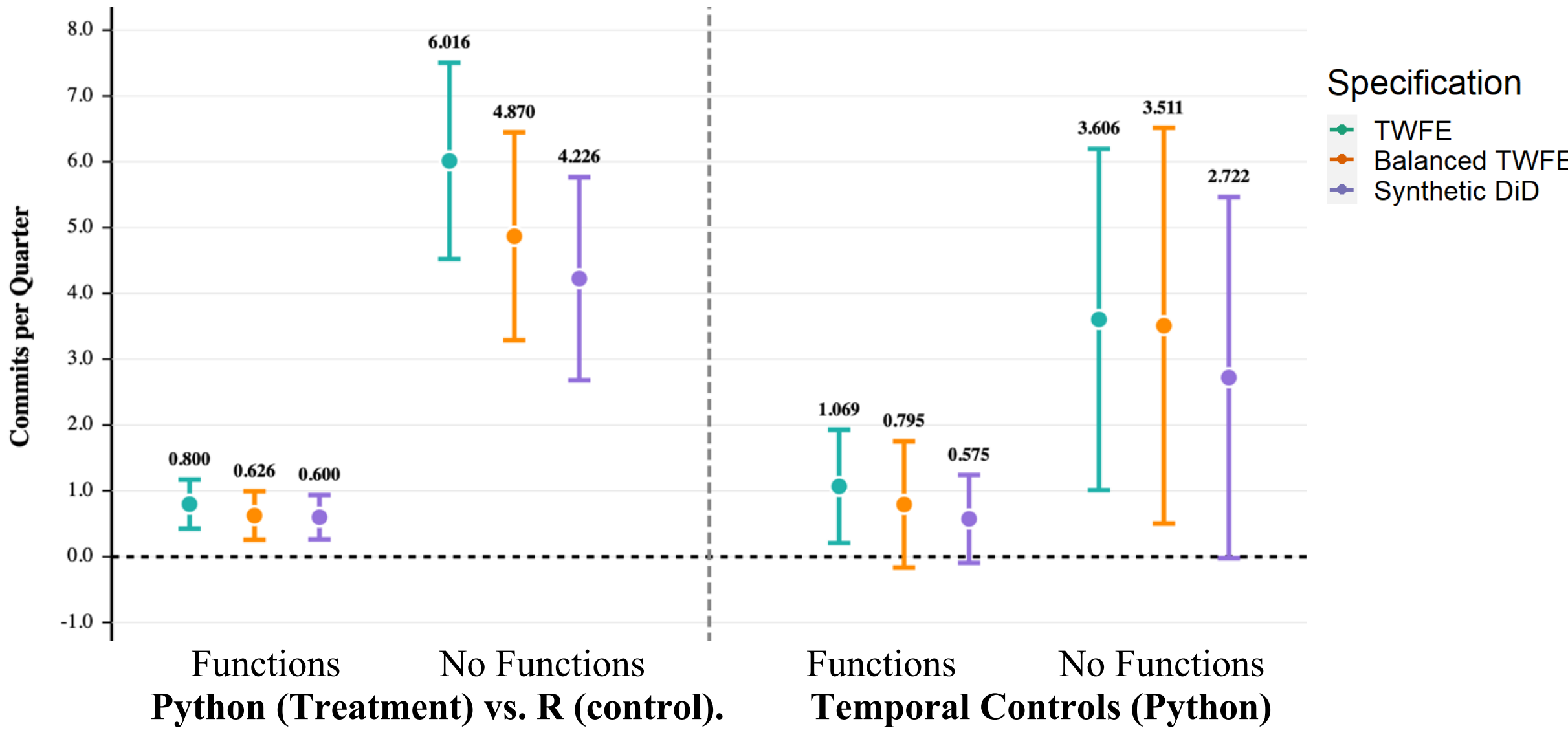

**Figure 7.** The Impact of GitHub Copilot on the Type of Contributions: Function Addition

### 5.2.2 LLM Classification: Code Development vs. Maintenance

Our second approach classifies commit messages into code development and maintenance categories using an LLM. Code development commits represent work that adds or extends functionality, while maintenance commits capture bug fixes, documentation updates, refactoring, and other upkeep activities.

Panel A of Table 5 reports results for Python versus R. For code development commits (Columns 1-3), GitHub Copilot increases such commits by 1.472 per package-quarter, with the balanced panel at 1.270 and synthetic DiD at 1.105, all significant at $p < 0.001$. For maintenance commits (Columns 4-6), the TWFE estimate is 2.299 additional commits per quarter, with the balanced panel at 2.023 and synthetic DiD at 1.873, again all at $p < 0.001$. As with the function analysis, both contribution types increase and the effect is larger for the more routine activity (maintenance). Triple-difference analyses confirm that these differences are statistically significant (Appendix Table A3). Appendix Figure A2 visualizes the Synthetic DiD estimation. Panel B of Table 5 presents the temporal controls analysis. The same ordering holds: maintenance increases more than code development. Figure 8 visualizes the outcomes for the 2 panels.

**Table 5. The Effect on the Type of Contributions: LLM Classification**

**Panel A: Python (Treatment) vs. R (Control) Analysis**

| | Code Development | | | Maintenance | | |
|---|---|---|---|---|---|---|
| Dependent Variable | (1) *TWFE* | (2) *Balanced TWFE* | (3) *SynthDiD* | (4) *TWFE* | (5) *Balanced TWFE* | (6) *SynthDiD* |
| $TreatmentLang_i$ X $afterCopilot_{it}$ | 1.472*** | 1.270*** | 1.105*** | 2.299*** | 2.023*** | 1.873*** |
| | (0.257) | (0.266) | (0.253) | (0.371) | (0.391) | (0.376) |
| Time Fixed Effect | YES | YES | YES | YES | YES | YES |
| Package Fixed Effect | YES | YES | YES | YES | YES | YES |
| # of Obs. | 29,050 | 24,739 | 24,739 | 29,050 | 24,739 | 24,739 |

**Panel B: Temporal Controls (Python)**

| | Code Development | | | Maintenance | | |
|---|---|---|---|---|---|---|
| Dependent Variable | (1) *TWFE* | (2) *Balanced TWFE* | (3) *SynthDiD* | (4) *TWFE* | (5) *Balanced TWFE* | (6) *SynthDiD* |
| $TreatmentLang_i$ X $afterCopilot_{it}$ | 1.404** | 1.351* | 0.835* | 2.349** | 2.356** | 2.010* |
| | (0.536) | (0.618) | (0.404) | (0.775) | (0.904) | (0.931) |
| Time Fixed Effect | YES | YES | YES | YES | YES | YES |
| Package Fixed Effect | YES | YES | YES | YES | YES | YES |
| # of Obs. | 22,677 | 19,426 | 19,426 | 22,677 | 19,426 | 19,426 |

Note: Standard Errors clustered around package in parentheses for columns (1), (2), (4) and (5). The variance for (3) and (6) are estimated using the Jackknife variance procedure.

*** p<0.001, ** p<0.01, * p<0.05, $^{+}$ p<0.1

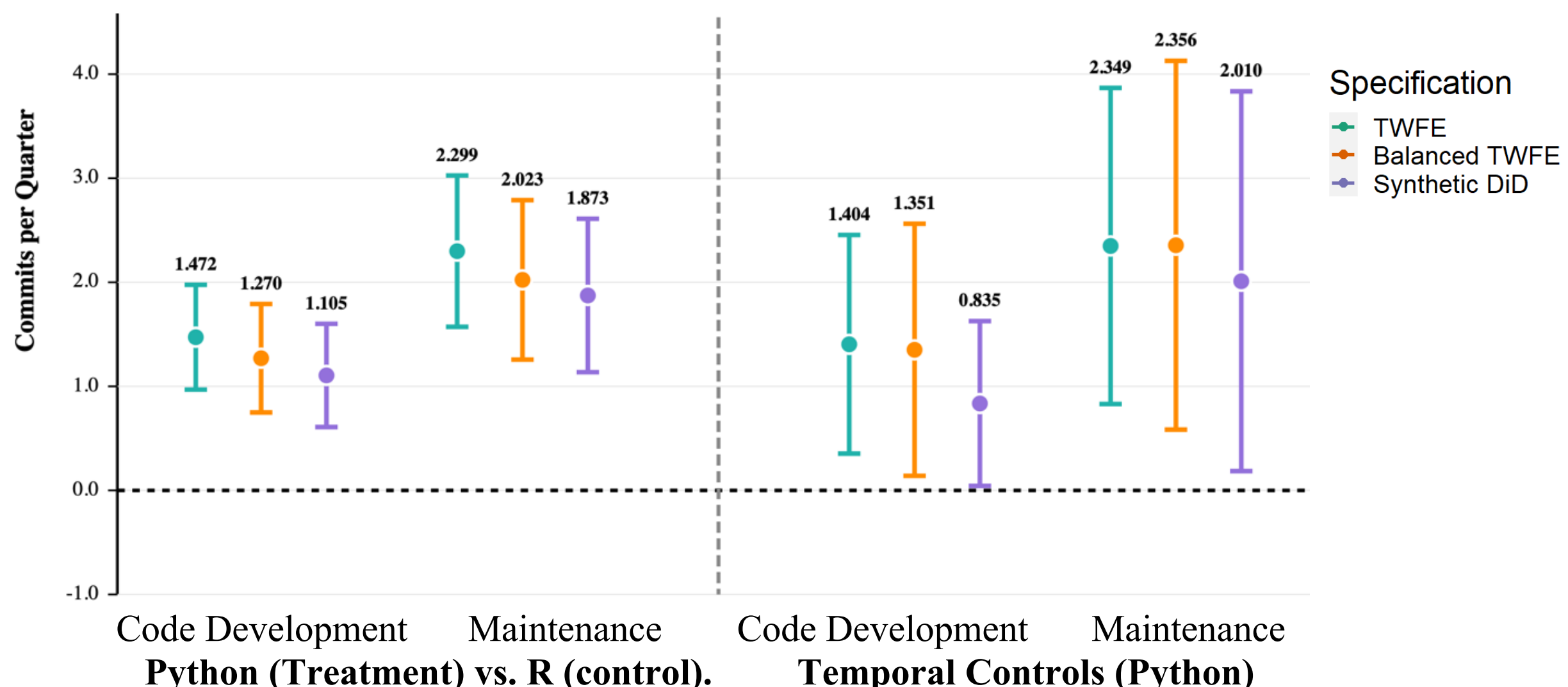

**Figure 8. The Impact of GitHub Copilot on the Type of Contributions: LLM Classification**

Analyses using both operationalizations of contributions (LLM Categorization and Function Detection) converge: Copilot increases both substantive and incremental contributions. However, incremental contributions increase more across all specifications, significantly more than the substantive contributions.

## 5.3 Heterogenous Effects based on Enhanced Ability to Provide/Utilize Contextual Content

The cross-language and temporal-controls analyses establish that Copilot affects contribution volume and composition. We next examine whether these effects depend on conditions that govern how effectively LLMs can deploy in-context learning. We focus on two such conditions: the richness of project-level context available to the model (Section 5.3.1) and the underlying capability of the model itself (Section 5.3.2). Both analyses use the Python-versus-R sample.

### 5.3.1 Project Activity Analysis

Projects with higher activity typically have more extensive code infrastructure, documentation, and discussion threads, providing richer context for LLM-based code generation. To investigate this heterogeneity, we perform a triple-difference analysis that interacts the DiD terms with $highProjectActivity_{i}$, an indicator for above-median pre-treatment commit volume, separately for each contribution type. Table 6 presents these results for Python versus R. In the function addition analysis, for low-activity projects, commits with new function definitions increase by 0.276 per quarter, while commits without new function definitions increase by 1.915 per quarter. For high-activity projects, the triple interaction terms reveal additional effects of 0.902 for function-adding commits and 7.942 for non-function-adding commits, yielding total effects in high-activity projects of 1.178 and 9.857, respectively.

The LLM categorization analysis shows a consistent pattern. For low-activity projects, code development commits increase by 0.254 per quarter and maintenance commits increase by 0.676. For high-activity projects, the triple interaction terms indicate additional effects of 2.342 for code development and 3.092 for maintenance, yielding total effects of 2.596 and 3.768. The gap between maintenance and code development is nearly three times larger in high-activity projects than in low-activity projects.

These results suggest that Copilot's effectiveness scales with the richness of available context. Projects with high activity have more extensive code infrastructure, discussion threads, and documentation, providing richer input for in-context learning. While contributions of both types increase in high-activity projects, the effects are larger for maintenance and non-function-adding commits, consistent with the theoretical mechanism that well-defined solution spaces benefit more from rich contextual information.

**Table 6. Heterogeneous Effect Based on Project Activity (Python vs. R)**

| | Function Addition Analysis | | | | LLM Categorization Analysis | | | |
|---|---|---|---|---|---|---|---|---|
| | New Functions Added | | New Functions Not Added | | Code Development | | Maintenance | |
| Dependent Variable | (1) *TWFE* | (2) *Balanced TWFE* | (3) *TWFE* | (4) *Balanced TWFE* | (5) *TWFE* | (6) *Balanced TWFE* | (7) *TWFE* | (8) *Balanced TWFE* |
| $TreatmentLang_i$ *X* | 0.276*** | 0.093 | 1.915*** | 1.279*** | 0.254*** | 0.086 | 0.676*** | 0.473** |
| $afterCopilot_{it}$ | (0.065) | (0.063) | (0.283) | (0.286) | (0.072) | (0.072) | (0.149) | (0.154) |
| $TreatmentLang_i$ *X* | 0.902* | 1.006* | 7.942*** | 7.241*** | 2.342*** | 2.387*** | 3.092*** | 3.106*** |
| *afterCopilot X* $highProjectActivity_i$ | (0.396) | (0.393) | (1.544) | (1.671) | (0.535) | (0.558) | (0.759) | (0.815) |
| Time Fixed Effect | YES | YES | YES | YES | YES | YES | YES | YES |
| Package Fixed Effect | YES | YES | YES | YES | YES | YES | YES | YES |
| # of Obs. | 29,050 | 24,739 | 29,050 | 24,739 | 29,050 | 24,739 | 29,050 | 24,739 |

Note: Standard Errors clustered around package in parentheses.
*** $p<0.001$, ** $p<0.01$, * $p<0.05$, $^{+}$ $p<0.1$

### 5.3.2 Model Upgrade Analysis

To examine how improvements in LLM capabilities affect the patterns documented above, we exploit the upgrade to GitHub Copilot's underlying model in June 2022.[9] We introduce an additional difference-in-differences term using a dummy variable for the post-upgrade period ($afterCopilotUpgrade_t$), estimating how the treatment effect changes with improved LLM capabilities.

[9] GitHub Copilot now has a better AI model and New Capabilities – Shuyin Zhao, GitHub Blog

Table 7 presents the results for Python versus R. In the function analysis, the initial Copilot release increases commit count with new function definitions by 0.593 per quarter and commits without new function definitions by 4.959. Following the model upgrade, we observe additional increases of 0.345 for function-adding commits and 1.762 for non-function-adding commits, indicating that the improvement benefits both contribution types while maintaining the gap between them. In the LLM categorization analysis, the initial release increases code development commits by 1.299 per quarter and maintenance commits by 1.730. Following the upgrade, the additional effect on code development is not statistically significant (0.288), while maintenance sees a statistically significant further increase (0.948).

Both analyses in this section point in the same direction: as LLMs improve their ability to process context, their impact on open-source contributions grows, and the differential between substantive and incremental contributions persists or widens. The model upgrade increases effects for both types, but the additional boost is larger for maintenance and non-function-adding contributions. This suggests that as LLM capabilities continue to advance, the gap between contribution types may continue to grow.

**Table 7. Effects of the GitHub Copilot Model Upgrade on Contribution Types (Python vs. R)**

| | Function Addition Analysis | | | | LLM Categorization Analysis | | | |
|---|---|---|---|---|---|---|---|---|
| | New Functions Added | | New Functions Not Added | | Code Development | | Maintenance | |
| Dependent Variable | (1) *TWFE* | (2) *Balanced TWFE* | (3) *TWFE* | (4) *Balanced TWFE* | (5) *TWFE* | (6) *Balanced TWFE* | (7) *TWFE* | (8) *Balanced TWFE* |
| $TreatmentLang_i$ X | 0.593** | 0.374 | 4.959*** | 3.503*** | 1.299*** | 1.021** | 1.730*** | 1.228** |
| $afterCopilot_{it}$ | (0.202) | (0.204) | (0.822) | (0.896) | (0.305) | (0.325) | (0.401) | (0.433) |
| $TreatmentLang_i$ X | 0.345* | 0.421* | 1.762* | 2.278** | 0.288 | $0.415^{+}$ | 0.948* | 1.325*** |
| $afterCopilotUpgrade_{it}$ | (0.157) | (0.164) | (0.717) | (0.785) | (0.229) | (0.250) | (0.375) | (0.399) |
| Time Fixed Effect | YES | YES | YES | YES | YES | YES | YES | YES |
| Package Fixed Effect | YES | YES | YES | YES | YES | YES | YES | YES |
| # of Obs. | 29,050 | 24,739 | 29,050 | 24,739 | 29,050 | 24,739 | 29,050 | 24,739 |

Note: Standard Errors clustered around package in parentheses.
*** p<0.001, ** p<0.01, * p<0.05, $^{+}$ p<0.1

## 5.4 Robustness Checks

We perform several additional analyses to ensure the robustness of our findings across different model specifications and potential temporal variations.

### 5.4.1 *Alternative Language Pair: Rust versus Haskell*

Our primary identification compares Python (treated) to R (control), and our complementary strategy uses temporal controls within Python. Because Python is the dominant language for AI and data science, it is useful to examine whether similar patterns emerge outside data-oriented languages. We therefore replicate the full analysis using Rust (treated) and Haskell (control), two statically typed, systems-oriented languages with active package ecosystems (Cargo and Hackage, respectively). Rust was supported by GitHub Copilot at launch while Haskell was not, providing the same natural experiment in a structurally different developer community. Following the same propensity score matching procedure, we obtain a matched sample of 1,373 packages in each language (Appendix Table B1).

Appendix B reports the complete set of results. The headline patterns from Sections 5.1 through 5.3 replicate in this alternative pair. Appendix Table B2 and Figure B2 show that Copilot increases the total volume of contributions to Rust packages relative to Haskell. Appendix Tables B3 and B4 confirm that the increase is larger for incremental contributions (commits without new function definitions and maintenance-classified commits) than for substantive contributions, consistent with the Python versus R results. The heterogeneity analyses also replicate: effects are concentrated in high-activity projects (Appendix Table B5), and the June 2022 model upgrade amplifies the differential rather than narrowing it (Appendix Table B6). Appendix Tables B7 through B13 further report robustness checks for this pair (alternative dependent variable, logged specifications, and temporal variations), all of which are consistent with the main findings.

We also construct a pooled analysis that combines all four languages: Haskell, Python, R, and Rust (Appendix Tables B14 through B18; Figure B5). Beyond testing whether treatment effects differ across the two language pairs via triple-difference specifications, the pooled design [10] expands the donor pool available for synthetic DiD estimation, which can improve the quality of the synthetic control when the additional control units are comparable to the treated units (Abadie 2021). The consistency of these patterns across two structurally different language pairs, and in the pooled specification, strengthens the interpretation that Copilot's impact on the composition of contributions reflects a general property of LLM-assisted coding rather than a feature specific to the Python ecosystem.

[10] Since Haskell and R packages share the key property of being unsupported by GitHub Copilot, pooling them as control units is a natural extension.

#### 5.4.2 *Alternative Dependent Variable: Package Version Releases*

Instead of using commit counts, we use the count of package updates per quarter as the dependent variable and estimate equation (1). Appendix Table A4 presents the results. The coefficient of 0.094 in column (1) means that GitHub Copilot increases the number of new package releases by 10.04% at a mean value of 0.936 releases per quarter. The results in columns (2) and (3) are consistent: the coefficients reflect a 10.68%, and 12.82% increase in the number of new Python package versions after the release of Copilot, compared to their matched counterparts. Appendix Figure A3 visualizes the Synthetic DiD results. Appendix Table B7 presents consistent results for Rust versus Haskell.

#### 5.4.3 *Alternative Model Specifications*

First, we estimate models using the log-transformed dependent variable to address the skewed distribution. The results for two methods of categorizing contribution types (function addition and LLM classification) are presented in Appendix Tables A5 and A6, respectively. In the function addition analysis (Appendix Table A5), commits with a new function definition increase by 0.186, which translates to 20.43% (($e^{0.186}$ - 1) * 100 = 20.43%), whereas commits without a new function definition increase by 0.397, which translates to 48.73%. In the LLM categorization (Appendix Table A6), code development commits increase by 0.191 (21.05%) and maintenance commits by 0.259 (29.57%). Appendix Tables B8 and B9 report consistent results for Rust versus Haskell.

#### 5.4.4 *Temporal Variations: Dropping Initial Three Quarters (0, 1, and 2)*

To avoid selection from early adopters of GitHub Copilot that may result in unique usage patterns, we perform the analysis by dropping the initial 3 quarters when GitHub Copilot was in its technical preview period. We present the results of the first specification in Appendix Table A7 (Function Addition Analysis) and Appendix Table A8 (LLM categorization Analysis). The results remain consistent with our main findings: both substantive and incremental contributions increase following GitHub Copilot's introduction, with a larger effect on incremental contributions. The coefficients in these models are generally larger than in our main specification, suggesting an accelerating trend of contributions to the supported languages as GitHub Copilot gained broader adoption. Appendix Tables B10 and B11 show consistent patterns for Rust versus Haskell.

#### 5.4.5 *Temporal Variations: Moving the Treatment to June 2021*

While GitHub Copilot was made available via various IDEs in October 2021, the announcement happened in June 2021, with a selective set of programmers receiving early access. To ensure our results are not sensitive to the precise timing of treatment, we advance our treatment window by one quarter (to June 2021). Appendix Table A9 presents results for function analysis, and Appendix Table A10 presents the results for LLM categorization with the earlier treatment date. The differential impact on contribution types remains consistent regardless of whether we consider October or June 2021 as the treatment period. Appendix Tables B12 and B13 confirm the same pattern for Rust versus Haskell.

## 6. Discussion and Conclusion

### 6.1 Discussion

The speed at which generative AI exploded across the knowledge economy creates a fundamental challenge for causal inference: constructing a credible counterfactual in the field has become difficult. The near-instantaneous and pervasive adoption of large language models makes it hard to find comparable settings where one group is exposed and another is unexposed. Our paper identifies one such natural experiment, arising from a business decision by GitHub not to support languages such as R (and Haskell) despite its backend model, Codex, being capable of processing R (and Haskell) code. This product-scope decision created an exogenous treatment-control partition between otherwise comparable programming ecosystems. Exploiting the window from GitHub Copilot's launch until ChatGPT's public release (five quarters), we find that Copilot availability increases the volume of open-source contributions by 28% to 40% across specifications (Table 3), corroborated by a temporal controls strategy that compares the same Python repositories across treatment and control eras. The volume increase is substantial, yet our primary finding concerns the composition of these contributions and what the compositional shift reveals about LLM-driven innovation.

Having operationalized the distinction between substantive and incremental contributions through two complementary classification approaches (Sections 3.3 and 5.2), we now interpret what these patterns imply for open-source innovation more broadly. Both contribution types increase following Copilot's introduction, but incremental contributions account for a disproportionate share of the growth. In both the function analysis and the LLM classification, the increase in incremental contributions (commits without new functions; maintenance commits) is statistically significantly larger than the increase in substantive contributions (commits adding new functions; code development commits), and the triple-difference analysis confirms this differential. These patterns align with the in-context learning mechanism developed in Section 2.2: LLMs generate outputs by identifying patterns across existing text and code (Brown et al. 2020), and tasks embedded in richer contextual environments, where the existing codebase, prior commits, and issue discussions collectively constrain both the problem and potential solutions, are more amenable to LLM assistance. Incremental contributions are situated in precisely such environments, while substantive contributions require creative problem formulation that existing context alone cannot fully specify.

Two additional findings reinforce this interpretation and suggest that the compositional shift is structural rather than transitional. The disparity between contribution types is more pronounced in projects with higher activity levels (Section 5.3.1), precisely the environments where contextual information available to LLMs is most abundant. The upgrade to Copilot's underlying model in June 2022 increases effects for both contribution types, but the additional boost is concentrated in incremental contributions (Section 5.3.2), indicating that model improvements reinforce the differential rather than closing it. Viewed through the lens of innovation theory, LLMs function as a context-dependent amplifier: they accelerate the exploitation of established codebases and conventions (March 1991; Benner and Tushman 2003) while contributing comparatively less to the exploratory work that generates novel functionality. This is consistent with Agrawal et al.'s (2018, 2023) framing of AI as a prediction technology, where tasks whose solution space is well constrained by existing context benefit disproportionately. For open-source ecosystems, LLM adoption may tilt the trajectory of collaborative innovation toward maintenance and refinement, a pattern that may intensify as models improve, unless incentives for substantive feature creation are introduced.

### 6.2 Contributions and Implications

Following Gopal et al (2024)'s contribution framework for IS research, our work offers theoretical, empirical, and practical contributions.

***Theoretical.*** We develop a cognitive-input framework (Section 2.2) that distinguishes substantive from incremental innovation based on the structure of cognitive input each task demands, extending classical innovation typologies developed for firm-bounded settings (Henderson and Clark 1990; Gatignon et al. 2002) to voluntary, collaborative contexts where contributors self-select into tasks. Within this framework, we identify in-context learning (Brown et al. 2020) as the mechanism linking LLM capabilities to differential effects across innovation types, and our heterogeneous effects and model upgrade analyses provide direct evidence for this mechanism. These findings contribute to the exploitation-exploration literature (March 1991; Benner and Tushman 2003) by demonstrating that a general-purpose AI technology produces a structurally similar tilt toward exploitation in voluntary knowledge work, through a mechanism rooted in the architecture of LLMs rather than in organizational incentives or process management.

***Empirical.*** We provide causal evidence from a rare field setting where a production-grade LLM was introduced with differential coverage across comparable ecosystems. Three identification strategies (cross-language comparison of Python and R, temporal controls comparing the same repositories across eras, and a replication with Rust and Haskell) and three estimation approaches (TWFE, balanced panel, and synthetic difference-in-differences) triangulate the findings. Two independent classification methods, function definition analysis and LLM-based commit classification, converge on the same compositional pattern. Unlike earlier studies that document declining knowledge exchange on user-generated content platforms following LLM introduction (Burtch et al. 2024; Quinn and Gutt 2025), we document increased contributions in open-source communities, extending field evidence on AI's effects beyond occupation-level exposure assessments (Lou et al. 2023), guided task settings (Peng et al. 2023), or consumer-facing deployments (Sun et al. 2024) to unguided, voluntary collaboration.

***Practical.*** The compositional shift toward incremental contributions has implications for open-source platforms and the organizations that depend on them. In voluntary settings where contributors choose their focus, LLM availability may draw effort toward tasks with well-defined solution spaces and readily verifiable outcomes, a self-selection pattern that is less constrained in open-source than in organizational settings with assigned tasks. Without deliberate intervention, such as grants, recognition programs, or targeted challenges for substantive feature development, this trajectory may be self-reinforcing as models improve. Designing such interventions is itself non-trivial: prior work on monetary sponsorship in open source shows that external funding can reallocate contributor effort across activities in ways that are not always anticipated (Medappa et al. 2023). Organizations that encourage employee contributions to open-source projects should consider directing effort toward early-stage projects where contextual infrastructure is still developing, precisely because these are the settings where LLM assistance provides the least relative advantage and where human creativity matters most.

### 6.3 Limitations

Our findings are bounded by a specific time window (October 2021 to December 2022) when GitHub Copilot was the primary LLM-powered code-completion tool available. This temporal boundary was necessary to maintain clean identification before ChatGPT's introduction changed the landscape. Copilot during this period offered code completion rather than interactive chat capabilities, so our findings reflect initial adoption patterns of a code-completion tool. Whether more sophisticated LLMs with conversational and agentic capabilities replicate or modify these patterns is an open question. Future research could collaborate with platforms offering richer LLM interfaces, or with organizations that restrict LLM use in some teams, to examine this.

Not every contributor to Python packages necessarily used Copilot during our study period, meaning our estimates likely represent a lower bound of the true effect. As adoption increases, we would expect larger effects on contribution patterns. Collaboration with GitHub to obtain sign-up data at the repository level would allow researchers to estimate an upper bound. Generalizing these findings beyond open-source also requires caution: the self-selection and contributor autonomy that characterize voluntary collaboration may produce different dynamics than organizational settings with formal task assignment.

While the in-context learning mechanism we develop in Section 2.2 is consistent with the empirical patterns we observe, including the heterogeneous effects and model upgrade results, we cannot directly observe the cognitive processes of individual developers as they interact with LLM tools. Whether the differential arises primarily because LLMs reduce the cognitive burden of comprehending existing code (as in pair programming; Cockburn and Williams 2000), because developers self-select into tasks where LLM assistance is more effective, or through some combination of these channels, remains an open question for future work that could combine observational data with experimental or survey-based approaches. Prior work shows that network centrality and value congruence independently shape which contributors remain active and what they work on (Maruping et al. 2019), suggesting that Copilot's effects may vary across contributor types, which is a dimension our package-level data cannot capture.

These limitations notwithstanding, our study provides field-level causal evidence on a question, how LLMs reshape innovation in collaborative settings, where such evidence remains scarce.

# Appendix A

## Appendix Figures

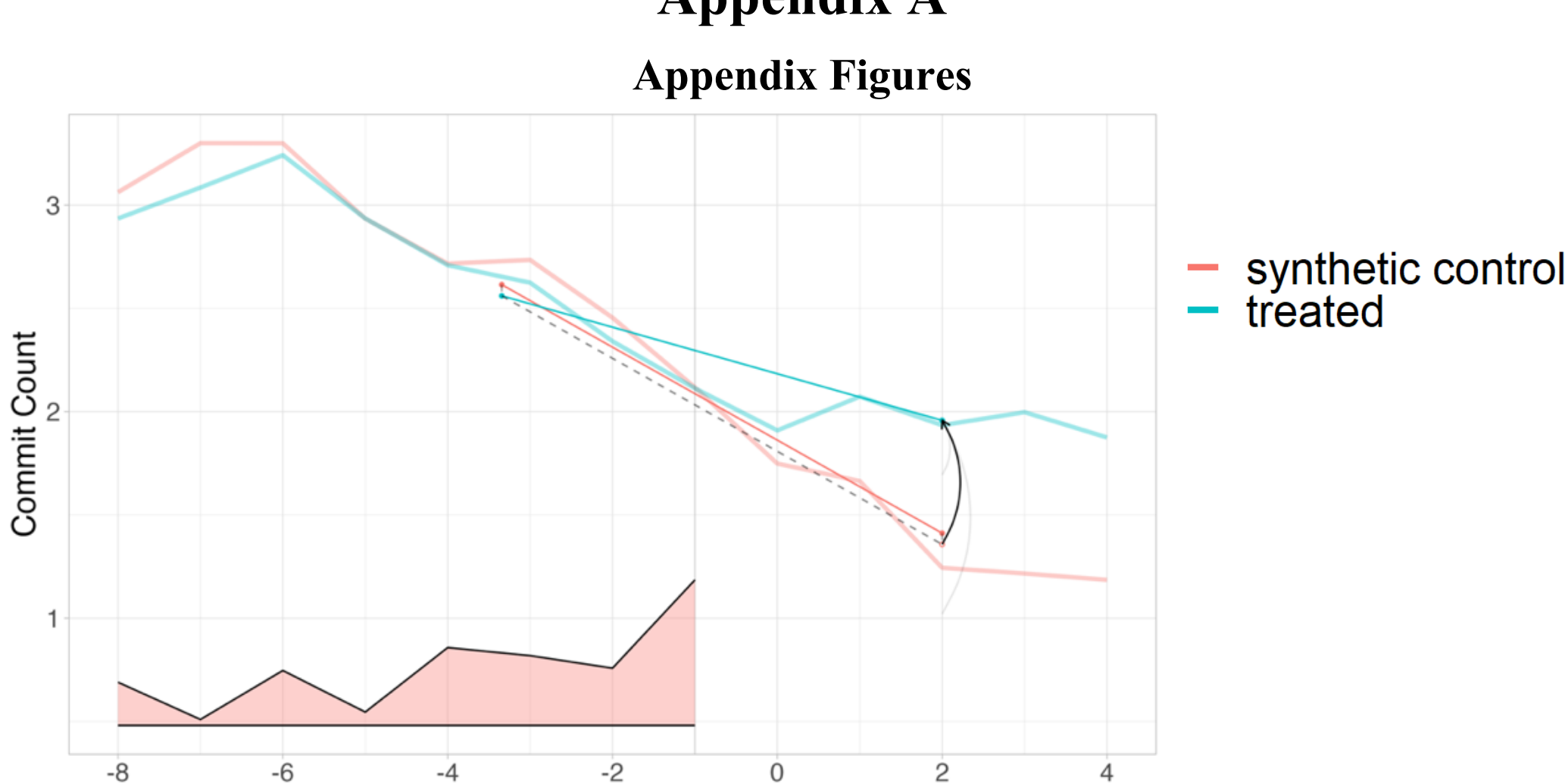

**Figure A1a.** The impact of Copilot on commits with new function definition (Python vs. R)

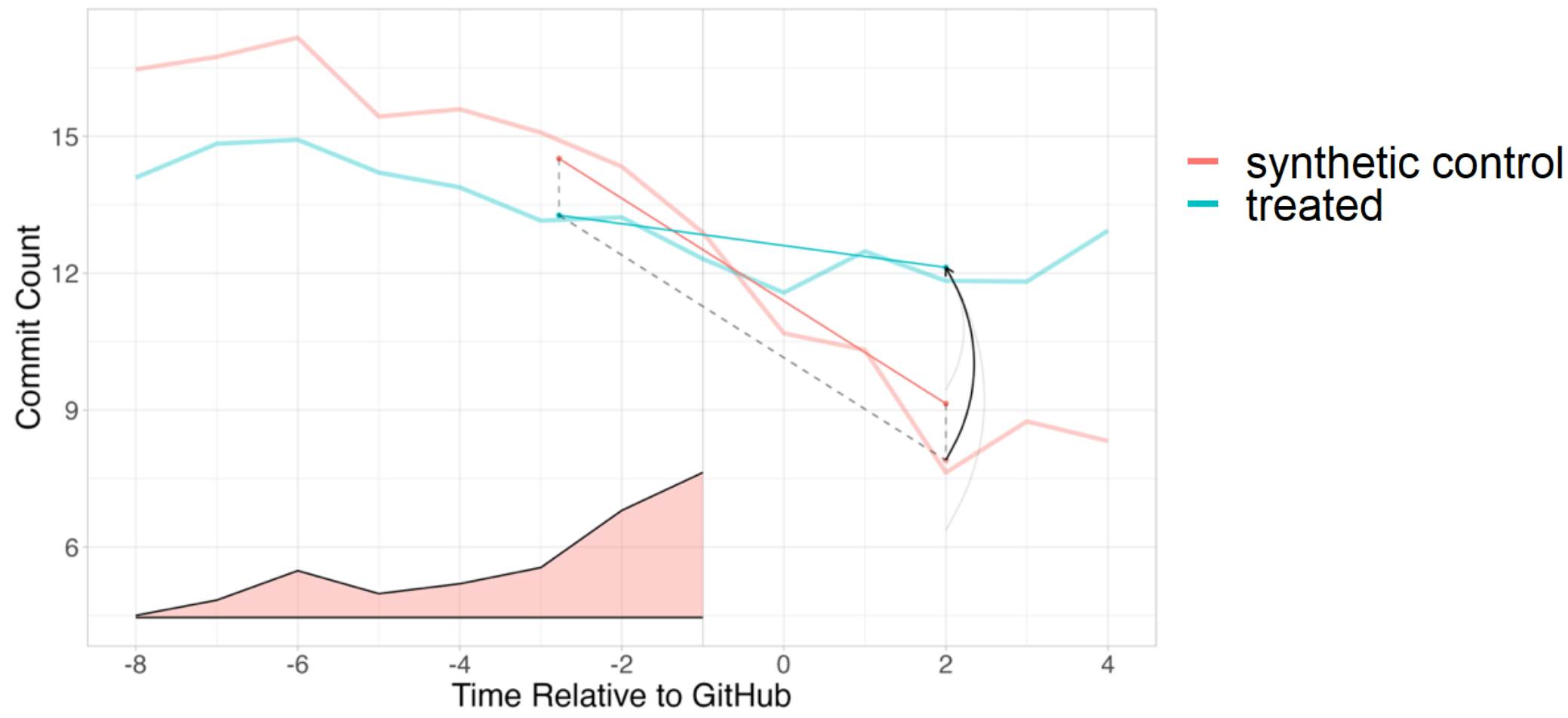

**Figure A1b.** The impact of Copilot on commits without new function definition (Python vs. R)

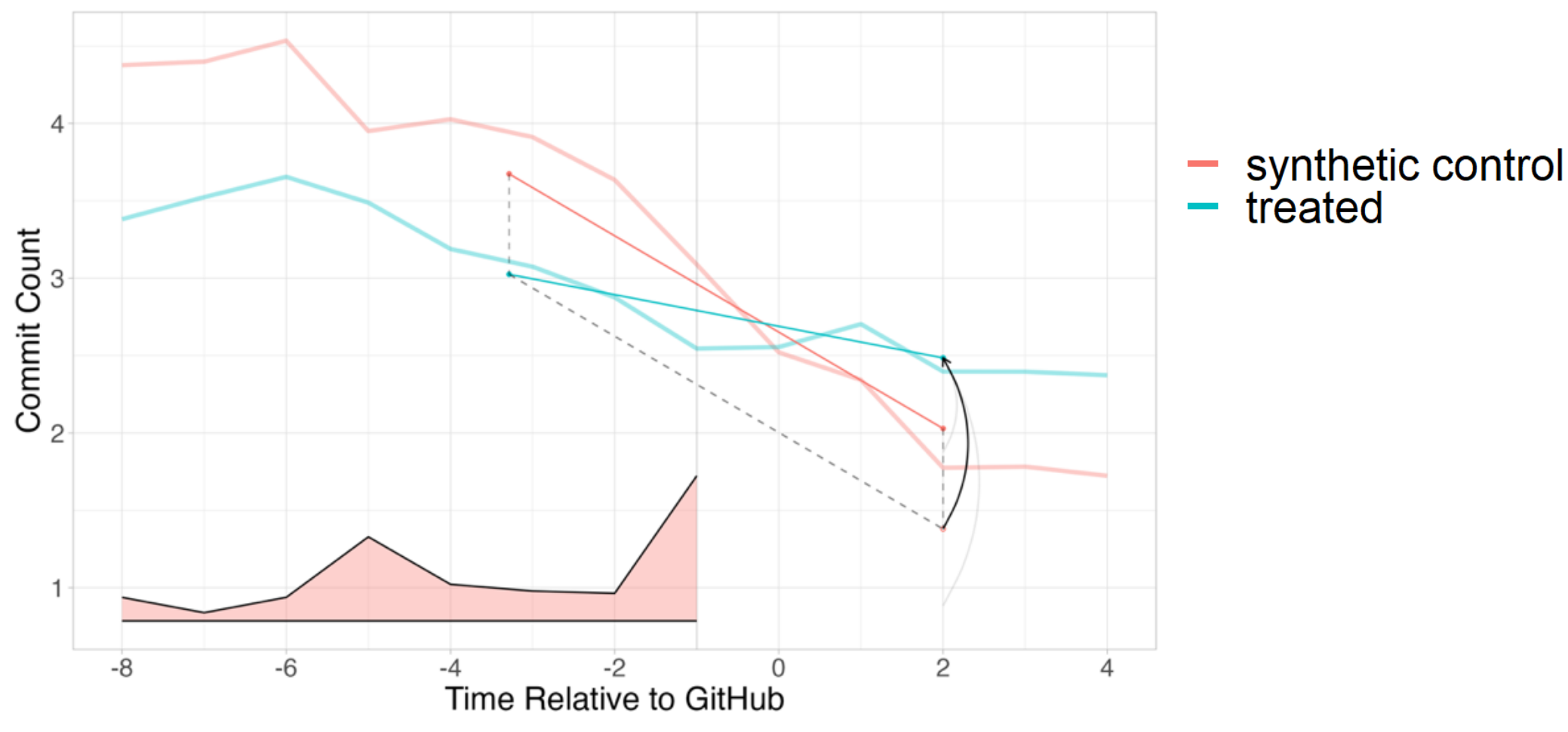

**Figure A2a.** The impact of Copilot on the count of *Code Development* commits (Python vs. R)

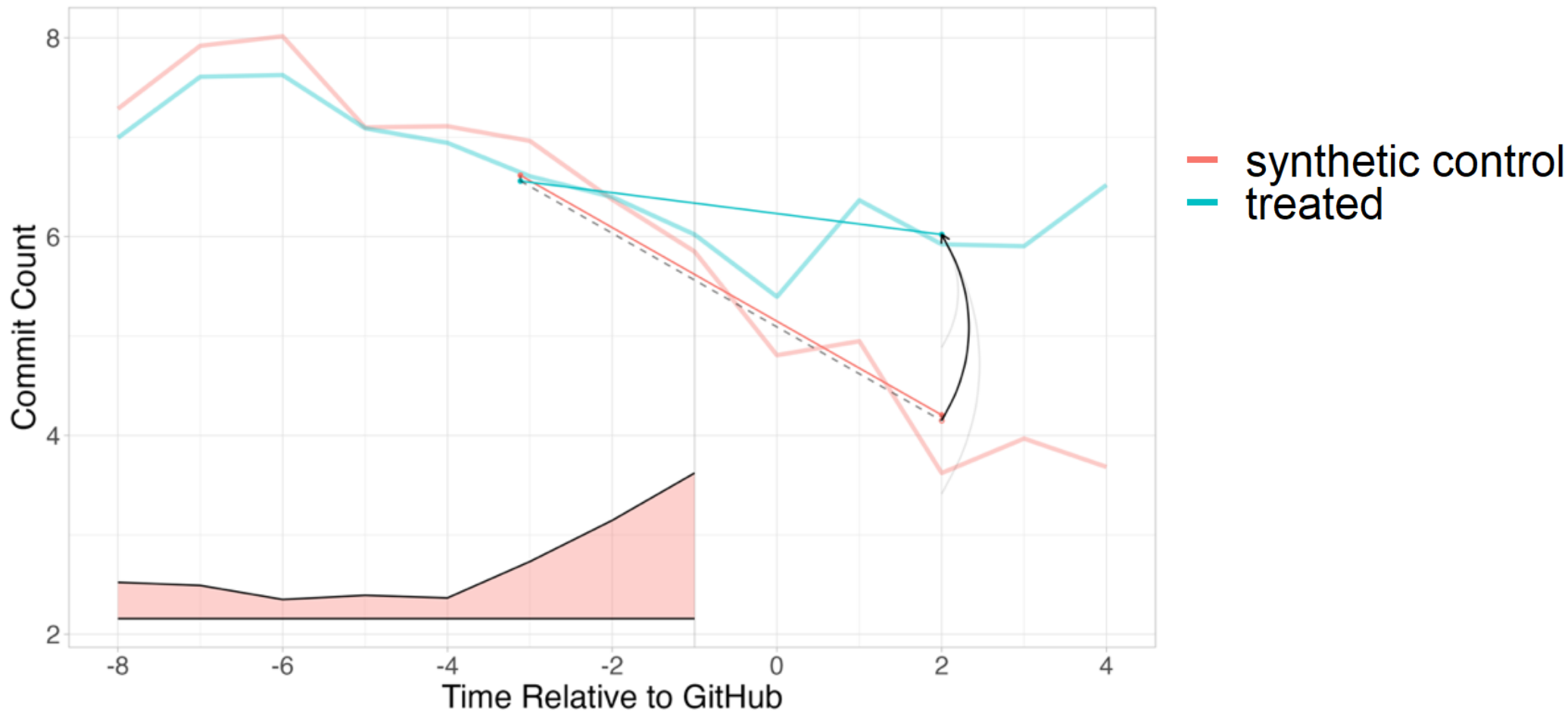

**Figure A2b.** The impact of Copilot on the count of *Maintenance* commits (Python vs. R)

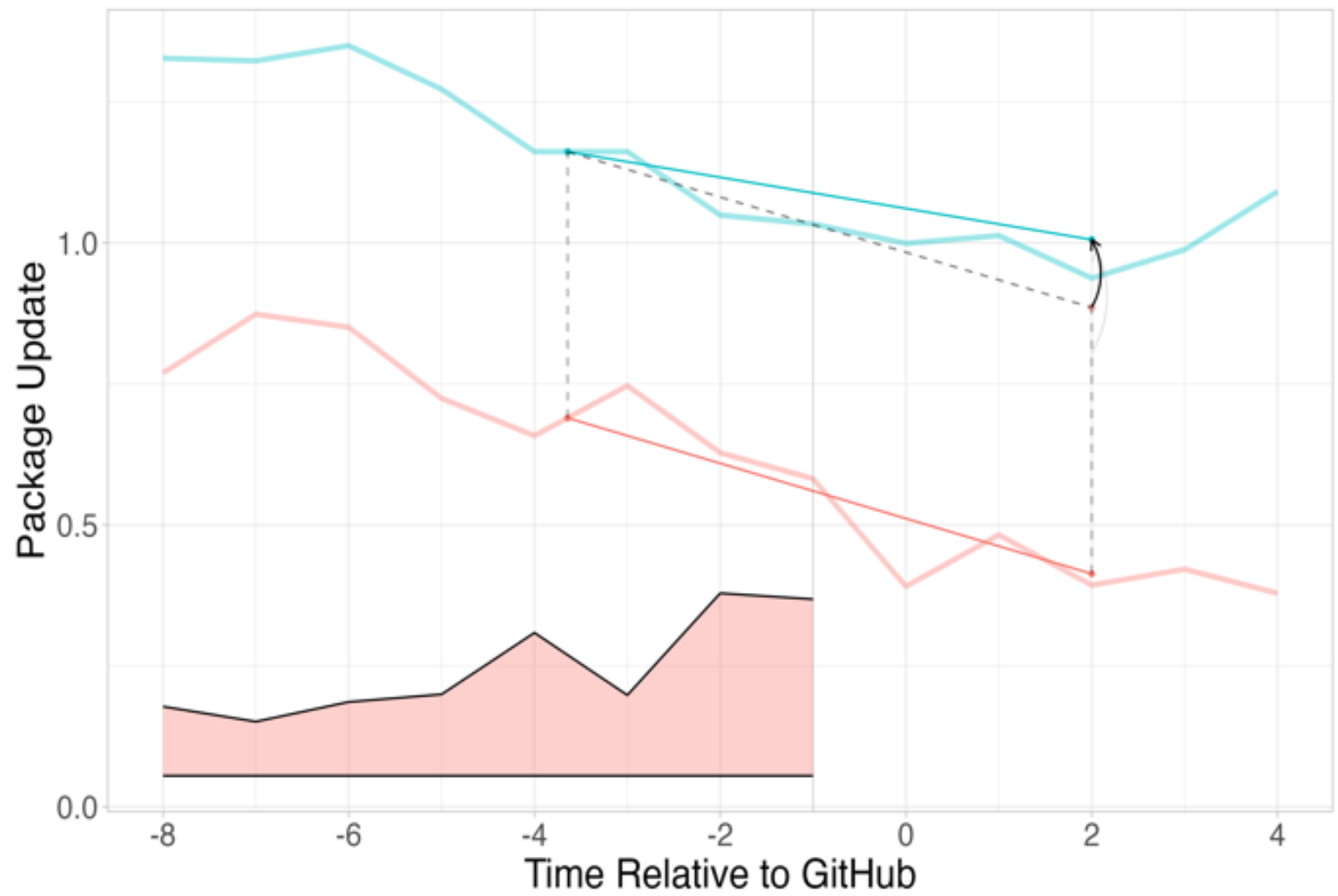

**Figure A3.** The impact of Copilot on New Version Release (Python vs. R)

## Appendix Tables

### Table A1. Summary Statistics (Python and R)

| | 1187 packages in main analyses | | | | |
|---|---|---|---|---|---|
| **Variables** | **# Obs.** | **Mean** | **Std. Dev.** | **Min** | **Max** |
| $Commits_{it}$ | 29,050 | 15.91 | 38.27 | 0 | 783 |
| $Commits\ (j{=}NewFunctiont)_{ijt}$ | 29,050 | 2.51 | 8.49 | 0 | 441 |
| $Commits((j{=}NoNewFunction)_{ijt}$ | 29,050 | 13.40 | 32.16 | 0 | 679 |
| $Commits\ (j{=}CodeDevelopment)_{ijt}$ | 29,050 | 3.22 | 11.04 | 0 | 396 |
| $Commits((j{=}Maintenance)_{ijt}$ | 29,050 | 6.41 | 15.67 | 0 | 341 |
| $VersionRelease_{it}$ | 29,050 | 0.94 | 1.88 | 0 | 39 |

Note: The final matched sample has 1,187 Python and R packages each.

### Table A2. Triple Difference: Function Addition Analysis (Python vs. R)

| Dependent Variable | Number of New Commits | |
|---|---|---|
| Models | (1)<br>*TWFE* | (2)<br>*Balanced TWFE* |
| $TreatmentLang_i \times afterCopilot_t$ | 0.919***<br>(0.206) | 0.626***<br>(0.188) |
| $TreatmentLang_i \times afterCopilot_t$<br>$\times\ Category_NoNewFunction_j$ | 4.979***<br>(0.653) | 4.244***<br>(0.697) |
| Time Fixed Effect | YES | YES |
| Package Fixed Effect | YES | YES |
| # of Obs. | 58,100 | 49,478 |

Note: Standard Errors clustered around package in parentheses
*** p<0.001, ** p<0.01, * p<0.05, + p<0.1

### Table A3. Triple Difference: LLM Categorization Analysis (Python vs. R)

| Dependent Variable | Number of New Commits | |
|---|---|---|
| Models | (1)<br>*TWFE* | (2)<br>*Balanced TWFE* |
| $TreatmentLang_i \times afterCopilot_t$ | 1.474***<br>(0.258) | 1.270***<br>(0.266) |
| $TreatmentLang_i \times afterCopilot_t$<br>$\times\ Category_Maintenance_j$ | 0.822**<br>(0.267) | 0.753**<br>(0.280) |
| Time Fixed Effect | YES | YES |
| Package Fixed Effect | YES | YES |
| # of Obs. | 58,100 | 49,478 |
| R-Squared | 0.508 | 0.512 |

Note: Standard Errors clustered around package in parentheses
*** p<0.001, ** p<0.01, * p<0.05, + p<0.1

### Table A4. The Effect on the Volume of Contributions: Alternative DV Measure (Python vs. R)

| Dependent Variable | Number of New Package Version Releases | | |
|---|---|---|---|
| Models | (1)<br>*TWFE* | (2)<br>*Balanced TWFE* | (3)<br>*SynthDiD* |
| $TreatmentLang_i \times afterCopilot_t$ | 0.094*<br>(0.046) | 0.100*<br>(0.044) | 0.120**<br>(0.041) |
| Time Fixed Effect | YES | YES | YES |
| Package Fixed Effect | YES | YES | YES |
| # of Obs. | 29,050 | 24,739 | 24,739 |

Note: *** p<0.001, ** p<0.01, * p<0.05
Standard Errors clustered around package in parentheses for (1) and (2).The variance for (3) is estimated using the Jackknife variance procedure.

**Table A5. The Effect on the Type of Contributions: Function Creation, Logged DV (Python vs. R)**

| Dependent Variable | Log(New Functions Added) | | | Log(New Functions Not Added) | | |
|---|---|---|---|---|---|---|
| | (1) *TWFE* | (2) *Balanced TWFE* | (3) *SynthDiD* | (4) *TWFE* | (5) *Balanced TWFE* | (6) *SynthDiD* |
| *$TreatmentLang_i$ X $afterCopilot_{it}$* | 0.186*** | 0.121*** | 0.109*** | 0.397*** | 0.310*** | 0.285*** |
| | (0.021) | (0.021) | (0.020) | (0.032) | (0.033) | (0.032) |
| Time Fixed Effect | YES | YES | YES | YES | YES | YES |
| Package Fixed Effect | YES | YES | YES | YES | YES | YES |
| # of Obs. | 29,050 | 24,739 | 24,739 | 29,050 | 24,739 | 24,739 |

Note: Standard Errors clustered around package in parentheses for columns (1), (2), (4) and (5). The variance for (3) and (6) are estimated using the Jackknife variance procedure.
*** $p<0.001$, ** $p<0.01$, * $p<0.05$, + $p<0.1$

**Table A6. The Effect on the Type of Contributions: LLM Classification, Logged DV (Python vs. R)**

| Dependent Variable | Log(Code Development) | | | Log(Maintenance) | | |
|---|---|---|---|---|---|---|
| | (1) *TWFE* | (2) *Balanced TWFE* | (3) *SynthDiD* | (4) *TWFE* | (5) *Balanced TWFE* | (6) *SynthDiD* |
| *$TreatmentLang_i$ X $afterCopilot_{it}$* | 0.191*** | 0.144*** | 0.130*** | 0.259*** | 0.208*** | 0.200*** |
| | (0.022) | (0.023) | (0.022) | (0.026) | (0.028) | (0.026) |
| Time Fixed Effect | YES | YES | YES | YES | YES | YES |
| Package Fixed Effect | YES | YES | YES | YES | YES | YES |
| # of Obs. | 29,050 | 24,739 | 24,739 | 29,050 | 24,739 | 24,739 |

Note: Standard Errors clustered around package in parentheses for columns (1), (2), (4) and (5). The variance for (3) and (6) are estimated using the Jackknife variance procedure.
*** $p<0.001$, ** $p<0.01$, * $p<0.05$, + $p<0.1$

**Table A7. The Effect on the Type of Contributions: Drop Quarters 0, 1, and 2 (Python vs. R)**

| Dependent Variable | New Functions Added | | | New Functions Not Added | | |
|---|---|---|---|---|---|---|
| | (1) *TWFE* | (2) *Balanced TWFE* | (3) *SynthDiD* | (4) *TWFE* | (5) *Balanced TWFE* | (6) *SynthDiD* |
| *$TreatmentLang_i$ X $afterCopilot_{it}$* | 1.007*** | 0.805*** | 0.805*** | 6.844*** | 5.689*** | 5.107*** |
| | (0.224) | (0.218) | (0.219) | (0.976) | (1.004) | (1.052) |
| Time Fixed Effect | YES | YES | YES | YES | YES | YES |
| Package Fixed Effect | YES | YES | YES | YES | YES | YES |
| # of Obs. | 21,928 | 19,030 | 19,030 | 21,928 | 19,030 | 19,030 |

Note: Standard Errors clustered around package in parentheses for columns (1), (2), (4) and (5). The variance for (3) and (6) are estimated using the Jackknife variance procedure.
*** $p<0.001$, ** $p<0.01$, * $p<0.05$, + $p<0.1$

**Table A8. The Effect on the Type of Contributions: Drop Quarters 0, 1, and 2 (Python vs. R)**

| Dependent Variable | Code Development | | | Maintenance | | |
|---|---|---|---|---|---|---|
| | (1) *TWFE* | (2) *Balanced TWFE* | (3) *SynthDiD* | (4) *TWFE* | (5) *Balanced TWFE* | (6) *SynthDiD* |
| *$TreatmentLang_i$ X $afterCopilot_{it}$* | 1.637*** | 1.422*** | 1.309*** | 2.732*** | 2.581*** | 2.447*** |
| | (0.287) | (0.290) | (0.299) | (0.482) | (0.484) | (0.497) |
| Time Fixed Effect | YES | YES | YES | YES | YES | YES |
| Package Fixed Effect | YES | YES | YES | YES | YES | YES |
| # of Obs. | 21,928 | 19,030 | 19,030 | 21,928 | 19,030 | 19,030 |

Note: Standard Errors clustered around package in parentheses for columns (1), (2), (4) and (5). The variance for (3) and (6) are estimated using the Jackknife variance procedure.
*** $p<0.001$, ** $p<0.01$, * $p<0.05$, + $p<0.1$

**Table A9. The Effect on the Type of Contributions: Move Treatment to June 2021 (Python vs. R)**

| Dependent Variable | New Functions Added | | | New Functions Not Added | | |
|---|---|---|---|---|---|---|
| | (1) *TWFE* | (2) *Balanced TWFE* | (3) *SynthDiD* | (4) *TWFE* | (5) *Balanced TWFE* | (6) *SynthDiD* |
| $TreatmentLang_i$ X $afterCopilot_{it}$ | 0.731*** | 0.553** | 0.559** | 5.650*** | 4.658*** | 4.192*** |
| | (0.204) | (0.199) | (0.182) | (0.779) | (0.813) | (0.793) |
| Time Fixed Effect | YES | YES | YES | YES | YES | YES |
| Package Fixed Effect | YES | YES | YES | YES | YES | YES |
| # of Obs. | 29,050 | 24,739 | 24,739 | 29,050 | 24,739 | 24,739 |

Note: Standard Errors clustered around package in parentheses for columns (1), (2), (4) and (5). The variance for (3) and (6) are estimated using the Jackknife variance procedure.
*** p<0.001, ** p<0.01, * p<0.05, + p<0.1

**Table A10. The Effect on the Type of Contributions: Move Treatment to June 2021 (Python vs. R)**

| Dependent Variable | Code Development | | | Maintenance | | |
|---|---|---|---|---|---|---|
| | (1) *TWFE* | (2) *Balanced TWFE* | (3) *SynthDiD* | (4) *TWFE* | (5) *Balanced TWFE* | (6) *SynthDiD* |
| $TreatmentLang_i$ X $afterCopilot_{it}$ | 1.414*** | 1.184*** | 1.059*** | 2.170*** | 1.865*** | 1.755*** |
| | (0.267) | (0.273) | (0.252) | (0.380) | (0.395) | (0.374) |
| Time Fixed Effect | YES | YES | YES | YES | YES | YES |
| Package Fixed Effect | YES | YES | YES | YES | YES | YES |
| # of Obs. | 29,050 | 24,739 | 24,739 | 29,050 | 24,739 | 24,739 |

Note: Standard Errors clustered around package in parentheses for columns (1), (2), (4) and (5). The variance for (3) and (6) are estimated using the Jackknife variance procedure.
*** p<0.001, ** p<0.01, * p<0.05, + p<0.1

# Appendix B

**Table B1. Summary Statistics (Haskell and Rust)**

| Variables | # Obs. | Mean | Std. Dev. | Min | Max |
|---|---|---|---|---|---|
| $Commits_{it}$ | 33,685 | 8.90 | 38.18 | 0 | 2329 |
| $Commits\ (j{=}NewFunction)_{ijt}$ | 33,685 | 2.34 | 11.21 | 0 | 382 |
| $Commits((j{=}NoNewFunction)_{ijt}$ | 33,685 | 6.56 | 28.37 | 0 | 1970 |
| $Commits\ (j{=}CodeDevelopment)_{ijt}$ | 33,685 | 2.13 | 11.72 | 0 | 774 |
| $Commits((j{=}Maintenance)_{ijt}$ | 33,685 | 3.93 | 16.96 | 0 | 1253 |
| $VersionRelease_{it}$ | 33,685 | 0.184 | 0.972 | 0 | 36 |

Note: The final matched sample has 1,373 Rust and Haskell packages each

**Table B2. The Effect on the Volume of Contributions (Rust vs. Haskell)**

| Dependent Variable | Number of New Commits | | |
|---|---|---|---|
| Models | (1) *TWFE* | (2) *Balanced TWFE* | (3) *SynthDiD* |
| $TreatmentLang_i\ X\ afterCopilot_t$ | 5.583***<br>(1.081) | 4.062***<br>(1.165) | 3.458***<br>(0.977) |
| Time Fixed Effect | YES | YES | YES |
| Package Fixed Effect | YES | YES | YES |
| # of Obs. | 33,685 | 28,197 | 28,197 |

Note: Standard Errors clustered around package in parentheses for columns (1), (2).
The variance for (3) is estimated using the Jackknife variance procedure.
*** p<0.001, ** p<0.01, * p<0.05

**Table B3. The Effect on the Type of Contributions: Function Addition (Rust vs. Haskell)**

**Rust (Treatment) vs. Haskell (Control) Analysis**

| | New Functions Added | | | New Functions Not Added | | |
|---|---|---|---|---|---|---|
| Dependent Variable | (1) *TWFE* | (2) *Balanced TWFE* | (3) *SynthDiD* | (4) *TWFE* | (5) *Balanced TWFE* | (6) *SynthDiD* |
| $TreatmentLang_i\ X\ afterCopilot_{it}$ | 1.879***<br>(0.284) | 1.238***<br>(0.274) | 0.859***<br>(0.212) | 3.704***<br>(0.841) | 2.824**<br>(0.932) | 2.621**<br>(0.805) |
| Time Fixed Effect | YES | YES | YES | YES | YES | YES |
| Package Fixed Effect | YES | YES | YES | YES | YES | YES |
| # of Obs. | 33,685 | 28,197 | 28,197 | 33,685 | 28,197 | 28,197 |

Note: Standard Errors clustered around package in parentheses for columns (1), (2), (4) and (5). The variance for (3) and (6) are estimated using the Jackknife variance procedure.
*** p<0.001, ** p<0.01, * p<0.05

**Table B4. The Effect on the Type of Contributions: LLM Classification (Rust vs. Haskell)**

**Rust (Treatment) vs. Haskell (Control) Analysis**

| | Code Development | | | Maintenance | | |
|---|---|---|---|---|---|---|
| Dependent Variable | (1) *TWFE* | (2) *Balanced TWFE* | (3) *SynthDiD* | (4) *TWFE* | (5) *Balanced TWFE* | (6) *SynthDiD* |
| $TreatmentLang_i\ X\ afterCopilot_{it}$ | 1.755***<br>(0.306) | 1.146***<br>(0.279) | 0.855***<br>(0.214) | 2.330***<br>(0.409) | 1.754***<br>(0.438) | 1.373***<br>(0.300) |
| Time Fixed Effect | YES | YES | YES | YES | YES | YES |
| Package Fixed Effect | YES | YES | YES | YES | YES | YES |
| # of Obs. | 33,685 | 28,197 | 28,197 | 33,685 | 28,197 | 28,197 |

Note: Standard Errors clustered around package in parentheses for columns (1), (2), (4) and (5). The variance for (3) and (6) are estimated using the Jackknife variance procedure.
*** p<0.001, ** p<0.01, * p<0.05, $^{+}$ p<0.1

**Table B5. Heterogeneous Effect Based on Project Activity (Rust vs. Haskell)**

| | Function Addition Analysis | | | | LLM Categorization Analysis | | | |
|---|---|---|---|---|---|---|---|---|
| | New Functions Added | | New Functions Not Added | | Code Development | | Maintenance | |
| Dependent Variable | (1) *TWFE* | (2) *Balanced TWFE* | (3) *TWFE* | (4) *Balanced TWFE* | (5) *TWFE* | (6) *Balanced TWFE* | (7) *TWFE* | (8) *Balanced TWFE* |
| $TreatmentLang_i$ X | -0.011 | 0.024 | 0.115 | 0.186 | -0.075 | 0.020 | 0.003 | 0.077 |
| $afterCopilot_{it}$ | (0.050) | (0.045) | (0.120) | (0.114) | (0.048) | (0.040) | (0.078) | (0.080) |
| $TreatmentLang_i$ X | 3.846*** | 2.724*** | 7.301*** | 5.918** | 2.878*** | 1.714** | 3.583*** | 2.504* |
| $highProjectActivity_i$ | (0.566) | (0.584) | (1.699) | (1.990) | (0.650) | (0.656) | (0.905) | (1.010) |
| Time Fixed Effect | YES | YES | YES | YES | YES | YES | YES | YES |
| Package Fixed Effect | YES | YES | YES | YES | YES | YES | YES | YES |
| # of Obs. | 33,685 | 28,197 | 33,685 | 28,197 | 33,685 | 28,197 | 33,685 | 28,197 |

Note: Standard Errors clustered around package in parentheses.
*** p<0.001, ** p<0.01, * p<0.05

**Table B6. Effects of the GitHub Copilot Model Upgrade (Rust vs. Haskell)**

| | Function Addition Analysis | | | | LLM Categorization Analysis | | | |
|---|---|---|---|---|---|---|---|---|
| Dependent Variable | New Functions Added | | New Functions Not Added | | Code Development | | Maintenance | |
| | (1) *TWFE* | (2) *Balanced TWFE* | (3) *TWFE* | (4) *Balanced TWFE* | (5) *TWFE* | (6) *Balanced TWFE* | (7) *TWFE* | (8) *Balanced TWFE* |
| $TreatmentLang_i$ X | 1.607*** | 1.056*** | 2.991*** | 2.288* | 1.432*** | 0.905** | 1.880*** | 1.423** |
| $afterCopilot_{it}$ | (0.298) | (0.296) | (0.828) | (0.926) | (0.306) | (0.283) | (0.433) | (0.480) |
| $TreatmentLang_i$ X | 0.453* | 0.303 | 1.188** | 0.892* | 0.538** | 0.402** | 0.751** | 0.552$^{+}$ |
| $afterCopilotUpgrade_{it}$ | (0.190) | (0.185) | (0.445) | (0.453) | (0.178) | (0.156) | (0.291) | (0.311) |
| Time Fixed Effect | YES | YES | YES | YES | YES | YES | YES | YES |
| Package Fixed Effect | YES | YES | YES | YES | YES | YES | YES | YES |
| # of Obs. | 33,685 | 28,197 | 33,685 | 28,197 | 33,685 | 28,197 | 33,685 | 28,197 |

Note: Standard Errors clustered around package in parentheses.
*** p<0.001, ** p<0.01, * p<0.05, $^{+}$ p<0.1

**Table B7. The Effect on the Volume of Contributions: Package Version Releases (Rust vs. Haskell)**

| Dependent Variable | New Package Version Release | | |
|---|---|---|---|
| Models | (4) *TWFE* | (5) *Balanced TWFE* | (6) *SynthDiD* |
| $TreatmentLang_i$ X $afterCopilot_t$ | 0.069*<br>(0.027) | 0.062*<br>(0.029) | 0.059*<br>(0.024) |
| Time Fixed Effect | YES | YES | YES |
| Package Fixed Effect | YES | YES | YES |
| # of Obs. | 33,685 | 28,197 | 28,197 |

Note: Standard Errors clustered around package in parentheses for columns (1), (2).
The variance for (3) is estimated using the Jackknife variance procedure.
*** p<0.001, ** p<0.01, * p<0.05

**Table B8. The Effect on the Type of Contributions: Logged DV (Rust vs. Haskell)**

| Dependent Variable | Log(New Functions Added) | | | Log(New Functions Not Added) | | |
|---|---|---|---|---|---|---|
| | (1) *TWFE* | (2) *Balanced TWFE* | (3) *SynthDiD* | (4) *TWFE* | (5) *Balanced TWFE* | (6) *SynthDiD* |
| *$TreatmentLang_i$ X $afterCopilot_{it}$* | 0.094*** | 0.074*** | 0.077*** | 0.099*** | 0.089*** | 0.099*** |
| | (0.019) | (0.019) | (0.018) | (0.025) | (0.026) | (0.026) |
| Time Fixed Effect | YES | YES | YES | YES | YES | YES |
| Package Fixed Effect | YES | YES | YES | YES | YES | YES |
| # of Obs. | 33,685 | 28,197 | 28,197 | 33,685 | 28,197 | 28,197 |

Note: Standard Errors clustered around package in parentheses for columns (1), (2), (4) and (5). The variance for (3) and (6) are estimated using the Jackknife variance procedure. *** p<0.001, ** p<0.01, * p<0.05, + p<0.1

**Table B9. The Effect on the Type of Contributions: Logged DV (Rust vs. Haskell)**

| Dependent Variable | Log(Code Development) | | | Log(Maintenance) | | |
|---|---|---|---|---|---|---|
| | (1) *TWFE* | (2) *Balanced TWFE* | (3) *SynthDiD* | (4) *TWFE* | (5) *Balanced TWFE* | (6) *SynthDiD* |
| *$TreatmentLang_i$ X $afterCopilot_{it}$* | 0.093*** | 0.073*** | 0.081*** | 0.114*** | 0.095*** | 0.093*** |
| | (0.018) | (0.018) | (0.017) | (0.021) | (0.023) | (0.024) |
| Time Fixed Effect | YES | YES | YES | YES | YES | YES |
| Package Fixed Effect | YES | YES | YES | YES | YES | YES |
| # of Obs. | 33,685 | 28,197 | 28,197 | 33,685 | 28,197 | 28,197 |

Note: Standard Errors clustered around package in parentheses for columns (1), (2), (4) and (5). The variance for (3) and (6) are estimated using the Jackknife variance procedure. *** p<0.001, ** p<0.01, * p<0.05, + p<0.1

**Table B10. The Effect on the Type of Contributions: Drop Quarters 0, 1, and 2 (Rust vs. Haskell)**

| Dependent Variable | New Functions Added | | | New Functions Not Added | | |
|---|---|---|---|---|---|---|
| | (1) *TWFE* | (2) *Balanced TWFE* | (3) *SynthDiD* | (4) *TWFE* | (5) *Balanced TWFE* | (6) *SynthDiD* |
| *$TreatmentLang_i$ X $afterCopilot_{it}$* | 2.225*** | 1.375*** | 1.041*** | 4.554*** | 3.245** | 2.852*** |
| | (0.330) | (0.305) | (0.247) | (0.923) | (0.990) | (0.862) |
| Time Fixed Effect | YES | YES | YES | YES | YES | YES |
| Package Fixed Effect | YES | YES | YES | YES | YES | YES |
| # of Obs. | 25,447 | 21,690 | 21,690 | 25,447 | 21,690 | 21,690 |

Note: Standard Errors clustered around package in parentheses for columns (1), (2), (4) and (5). The variance for (3) and (6) are estimated using the Jackknife variance procedure. *** p<0.001, ** p<0.01, * p<0.05, + p<0.1

**Table B11. The Effect on the Type of Contributions: Drop Quarters 0, 1, and 2 (Rust vs. Haskell)**

| Dependent Variable | Code Development | | | Maintenance | | |
|---|---|---|---|---|---|---|
| | (1) *TWFE* | (2) *Balanced TWFE* | (3) *SynthDiD* | (4) *TWFE* | (5) *Balanced TWFE* | (6) *SynthDiD* |
| *$TreatmentLang_i$ X $afterCopilot_{it}$* | 2.166*** | 1.324*** | 1.079*** | 2.812*** | 2.016*** | 1.551*** |
| | (0.384) | (0.317) | (0.267) | (0.464) | (0.492) | (0.382) |
| Time Fixed Effect | YES | YES | YES | YES | YES | YES |
| Package Fixed Effect | YES | YES | YES | YES | YES | YES |
| # of Obs. | 25,447 | 21,690 | 21,690 | 25,447 | 21,690 | 21,690 |

Note: Standard Errors clustered around package in parentheses for columns (1), (2), (4) and (5). The variance for (3) and (6) are estimated using the Jackknife variance procedure. *** p<0.001, ** p<0.01, * p<0.05, + p<0.1

**Table B12. The Effect on the Type of Contributions: Move Treatment to June 2021 (Rust vs. Haskell)**

| | New Functions Added | | | New Functions Not Added | | |
|---|---|---|---|---|---|---|
| Dependent Variable | (1) *TWFE* | (2) *Balanced TWFE* | (3) *SynthDiD* | (4) *TWFE* | (5) *Balanced TWFE* | (6) *SynthDiD* |
| $TreatmentLang_i$ X $afterCopilot_{it}$ | 1.734*** | 1.238*** | 1.050*** | 3.309*** | 2.728** | 2.570*** |
| | (0.283) | (0.285) | (0.223) | (0.810) | (0.897) | (0.706) |
| Time Fixed Effect | YES | YES | YES | YES | YES | YES |
| Package Fixed Effect | YES | YES | YES | YES | YES | YES |
| # of Obs. | 33,685 | 28,197 | 28,197 | 33,685 | 28,197 | 28,197 |

Note: Standard Errors clustered around package in parentheses for columns (1), (2), (4) and (5). The variance for (3) and (6) are estimated using the Jackknife variance procedure.
*** p<0.001, ** p<0.01, * p<0.05, $^{+}$ p<0.1

**Table B13. The Effect on the Type of Contributions: Move Treatment to June 2021 (Rust vs. Haskell)**

| | Code Development | | | Maintenance | | |
|---|---|---|---|---|---|---|
| Dependent Variable | (1) *TWFE* | (2) *Balanced TWFE* | (3) *SynthDiD* | (4) *TWFE* | (5) *Balanced TWFE* | (6) *SynthDiD* |
| $TreatmentLang_i$ X $afterCopilot_{it}$ | 1.501*** | 1.136*** | 0.972*** | 2.193*** | 1.766*** | 1.444*** |
| | (0.282) | (0.296) | (0.218) | (0.436) | (0.468) | (0.286) |
| Time Fixed Effect | YES | YES | YES | YES | YES | YES |
| Package Fixed Effect | YES | YES | YES | YES | YES | YES |
| # of Obs. | 33,685 | 28,197 | 28,197 | 33,685 | 28,197 | 28,197 |

Note: Standard Errors clustered around package in parentheses for columns (1), (2), (4) and (5). The variance for (3) and (6) are estimated using the Jackknife variance procedure. *** p<0.001, ** p<0.01, * p<0.05, $^{+}$ p<0.1

**Table B14. Number of New Commits: Combined Language Pairs**

| Comparison | Python (Treated) vs. R (Control) | | |
|---|---|---|---|
| Models | (1) *TWFE* | (2) *Balanced TWFE* | (3) *SynthDiD* |
| $TreatmentLang_i$ X $afterCopilot_t$ | 6.120*** (0.712) | 4.563*** (0.748) | 4.058*** (0.683) |
| Time Fixed Effect | YES | YES | YES |
| Package Fixed Effect | YES | YES | YES |
| # of Obs. | 62,735 | 52,936 | 52,936 |

Note: Standard Errors clustered around package in parentheses for columns (1), (2), (4) and (5). The variance for (3) and (6) are estimated using the Jackknife variance procedure. *** p<0.001, ** p<0.01, * p<0.05

**Table B15. Triple Difference: Function Addition Analysis (Rust vs. Haskell)**

| Dependent Variable | Number of New Commits | |
|---|---|---|
| Models | (1) *TWFE* | (2) *Balanced TWFE* |
| $TreatmentLang_i$ X $afterCopilot_t$ | 1.839*** (0.300) | 1.238*** (0.274) |
| $TreatmentLang_i$ X $afterCopilot_t$ X $Category_NoNewFunction_j$ | 1.905** (0.618) | 1.586* (0.728) |
| Time Fixed Effect | YES | YES |
| Package Fixed Effect | YES | YES |
| # of Obs. | 67,370 | 56,394 |

Note: Standard Errors clustered around package in parentheses
*** p<0.001, ** p<0.01, * p<0.05, $^{+}$ p<0.1

**Table B16. The Effect on the Type of Contributions: All Languages**
**Python + Rust (Treatment) vs. R + Haskell (Control): New Function Analysis**

| | New Functions Added | | | New Functions Not Added | | |
|---|---|---|---|---|---|---|
| Dependent Variable | (1) *TWFE* | (2) *Balanced TWFE* | (3) *SynthDiD* | (4) *TWFE* | (5) *Balanced TWFE* | (6) *SynthDiD* |
| *$TreatmentLang_i$ X $afterCopilot_{it}$* | 1.389*** | 0.943*** | 0.807*** | 4.731*** | 3.620*** | 3.260*** |
| | (0.177) | (0.172) | (0.156) | (0.571) | (0.611) | (0.569) |
| Time Fixed Effect | YES | YES | YES | YES | YES | YES |
| Package Fixed Effect | YES | YES | YES | YES | YES | YES |
| # of Obs. | 62,735 | 52,936 | 52,936 | 62,735 | 52,936 | 52,936 |

Note: *** p<0.001, ** p<0.01, * p<0.05
Standard Errors clustered around package in parentheses for (1), (2), (4) and (5). The variance for (3) and (6) are estimated using the Jackknife variance procedure.

**Table B17. Triple Difference: LLM Classification Analysis (Rust vs. Haskell)**

| Dependent Variable | Number of New Commits | |
|---|---|---|
| Models | (1) *TWFE* | (2) *Balanced TWFE* |
| *$TreatmentLang_i$ X $afterCopilot_t$* | 1.704*** | 1.146*** |
| | (0.294) | (0.279) |
| *$TreatmentLang_i$ X $afterCopilot_t$ X $Category_Maintenance_j$* | 0.678** | 0.608** |
| | (0.222) | (0.230) |
| Time Fixed Effect | YES | YES |
| Package Fixed Effect | YES | YES |
| # of Obs. | 67,370 | 56,394 |
| R-Squared | 0.450 | 0.467 |

Note: Standard Errors clustered around package in parentheses
*** p<0.001, ** p<0.01, * p<0.05, + p<0.1

**Table B18. The Effect on the Type of Contributions: All Languages**
**Python + Rust (Treatment) vs. R + Haskell (Control): LLM Categorization**

| | Code Development | | | Maintenance | | |
|---|---|---|---|---|---|---|
| Dependent Variable | (1) *TWFE* | (2) *Balanced TWFE* | (3) *SynthDiD* | (4) *TWFE* | (5) *Balanced TWFE* | (6) *SynthDiD* |
| *$TreatmentLang_i$ X $afterCopilot_{it}$* | 1.621*** | 1.170*** | 0.994*** | 2.313*** | 1.838*** | 1.596*** |
| | (0.204) | (0.195) | (0.174) | (0.281) | (0.299) | (0.252) |
| Time Fixed Effect | YES | YES | YES | YES | YES | YES |
| Package Fixed Effect | YES | YES | YES | YES | YES | YES |
| # of Obs. | 62,735 | 52,936 | 52,936 | 62,735 | 52,936 | 52,936 |

Note: *** p<0.001, ** p<0.01, * p<0.05
Standard Errors clustered around package in parentheses for (1), (2), (4) and (5). The variance for (3) and (6) are estimated using the Jackknife variance procedure.

## Appendix B Figures

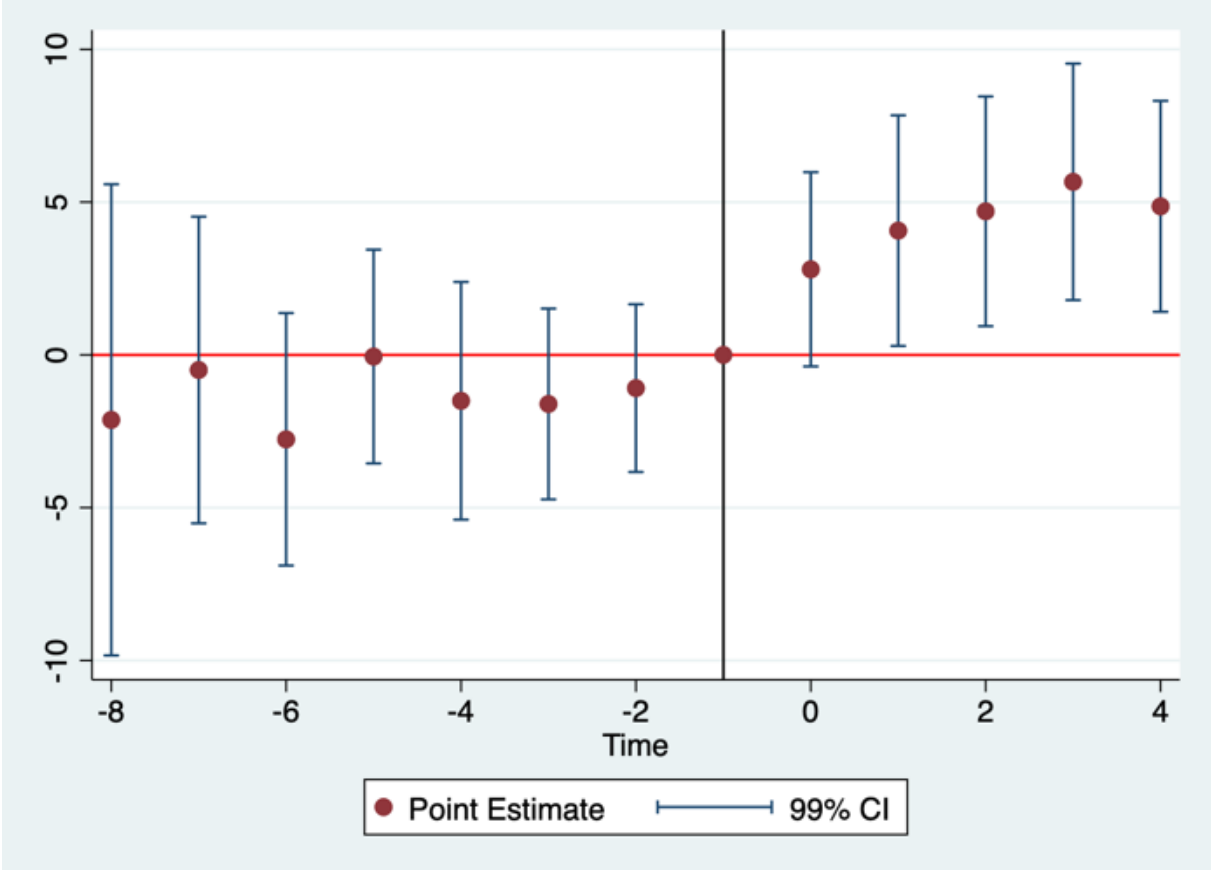

Rust vs. Haskell Parallel Trends

**Figure B1**. Parallel Trends Graph.

Note: The y-axis is the raw value of the commit count each quarter. 99% confidence interval bars.

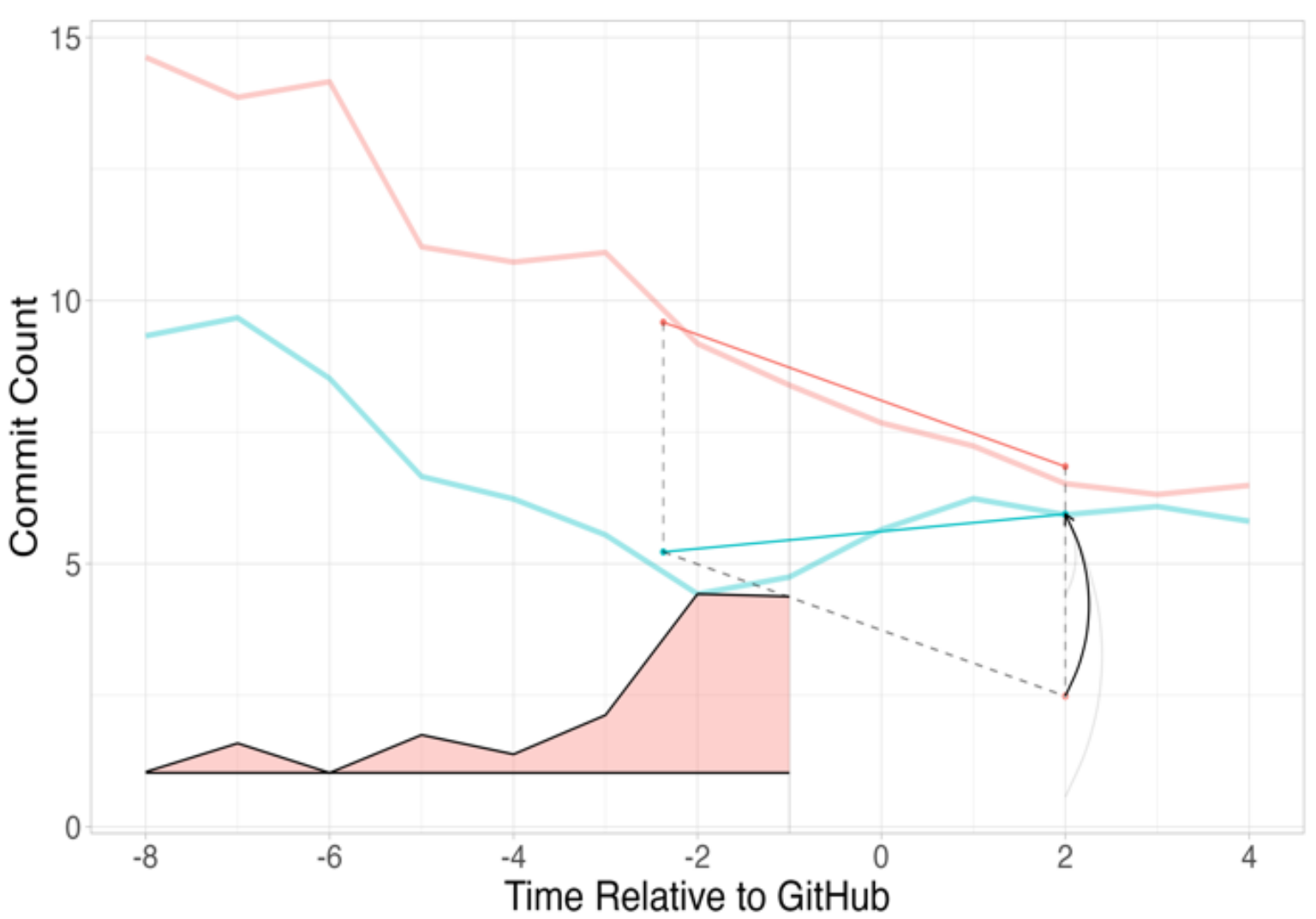

**Figure B2.** Synthetic DiD: Total Contributions (Rust vs. Haskell)

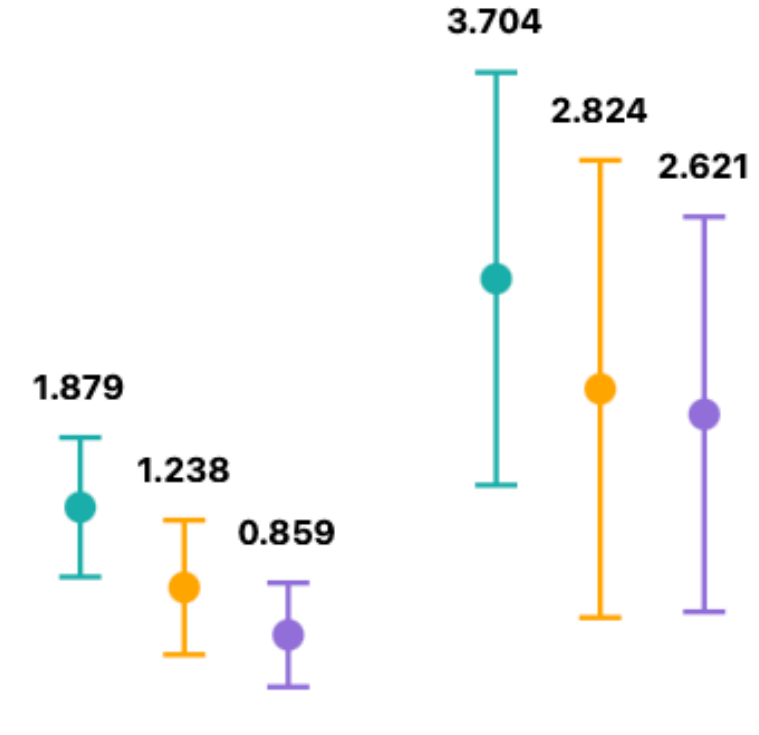

**Figure B3.** Synthetic DiD: Function Addition (Rust vs. Haskell)

Note: The y-axis is the raw value of the commit count each quarter. 99% confidence interval bars.

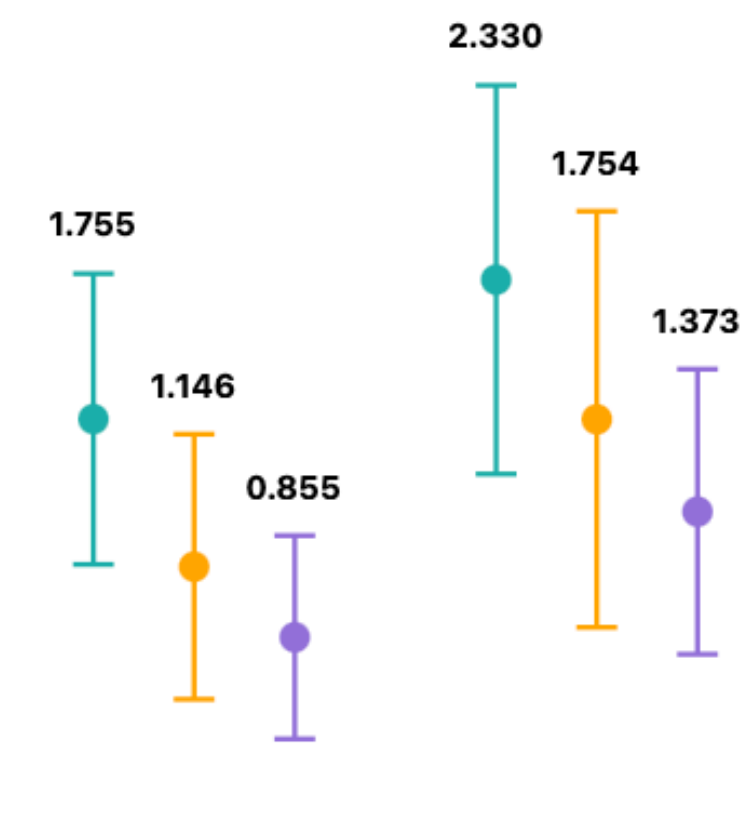

**Figure B4.** Synthetic DiD: LLM Classification (Rust vs. Haskell)

Note: The y-axis is the raw value of the commit count each quarter. 99% confidence interval bars.

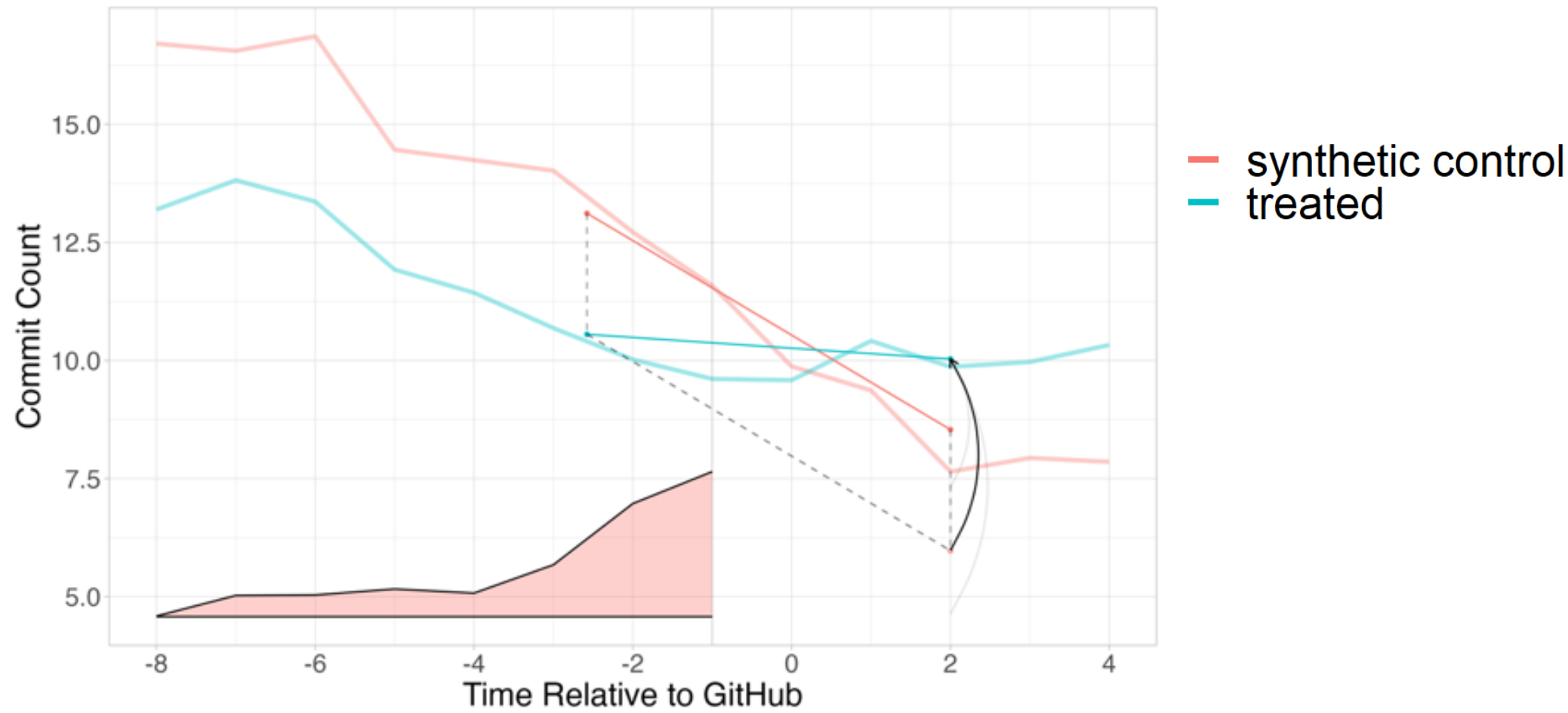

**Figure B5.** Treated (Python and Rust) versus Control (R and Haskell)

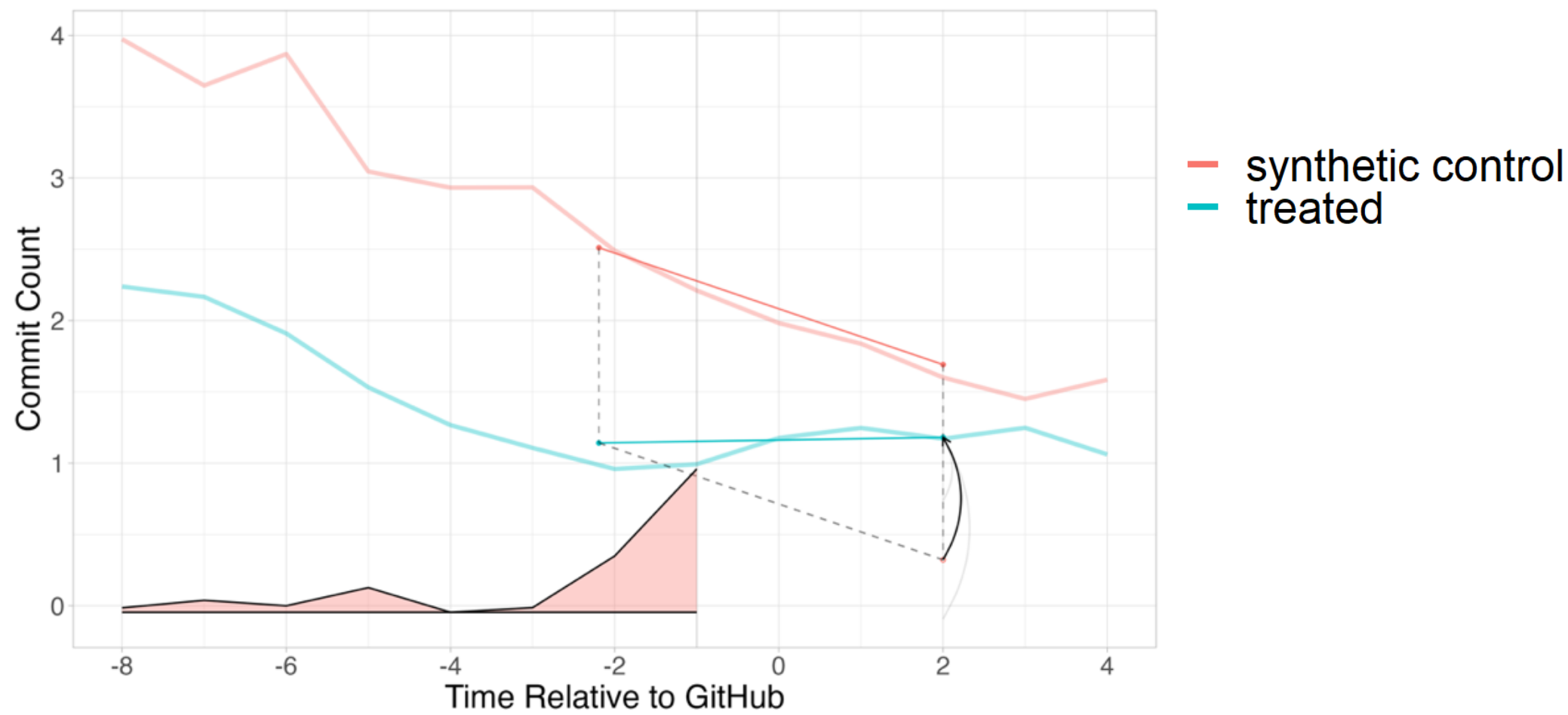

**Figure B6a.** The impact of Copilot on commits with new function definition (Rust vs. Haskell)

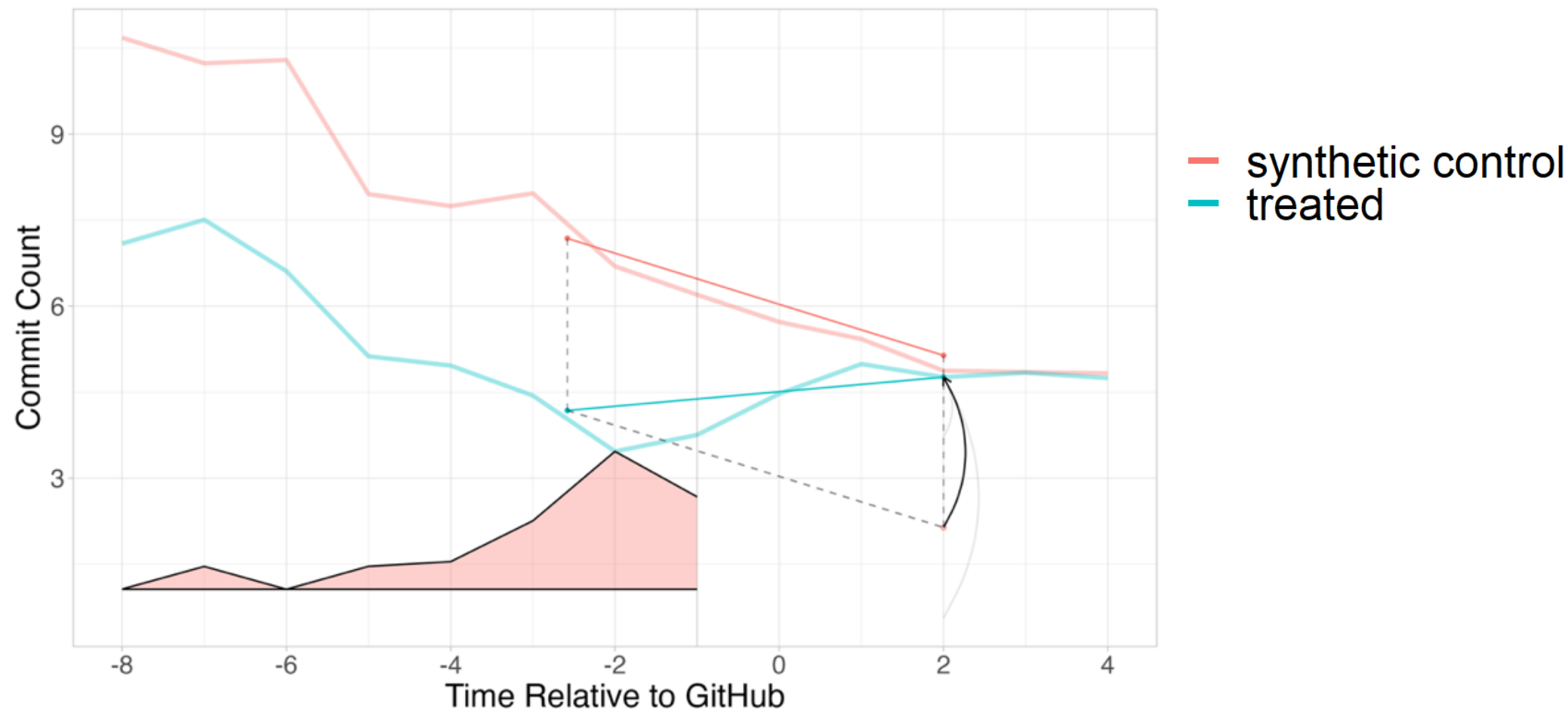

**Figure B6b.** The impact of Copilot on commits without new function definition (Rust vs. Haskell)

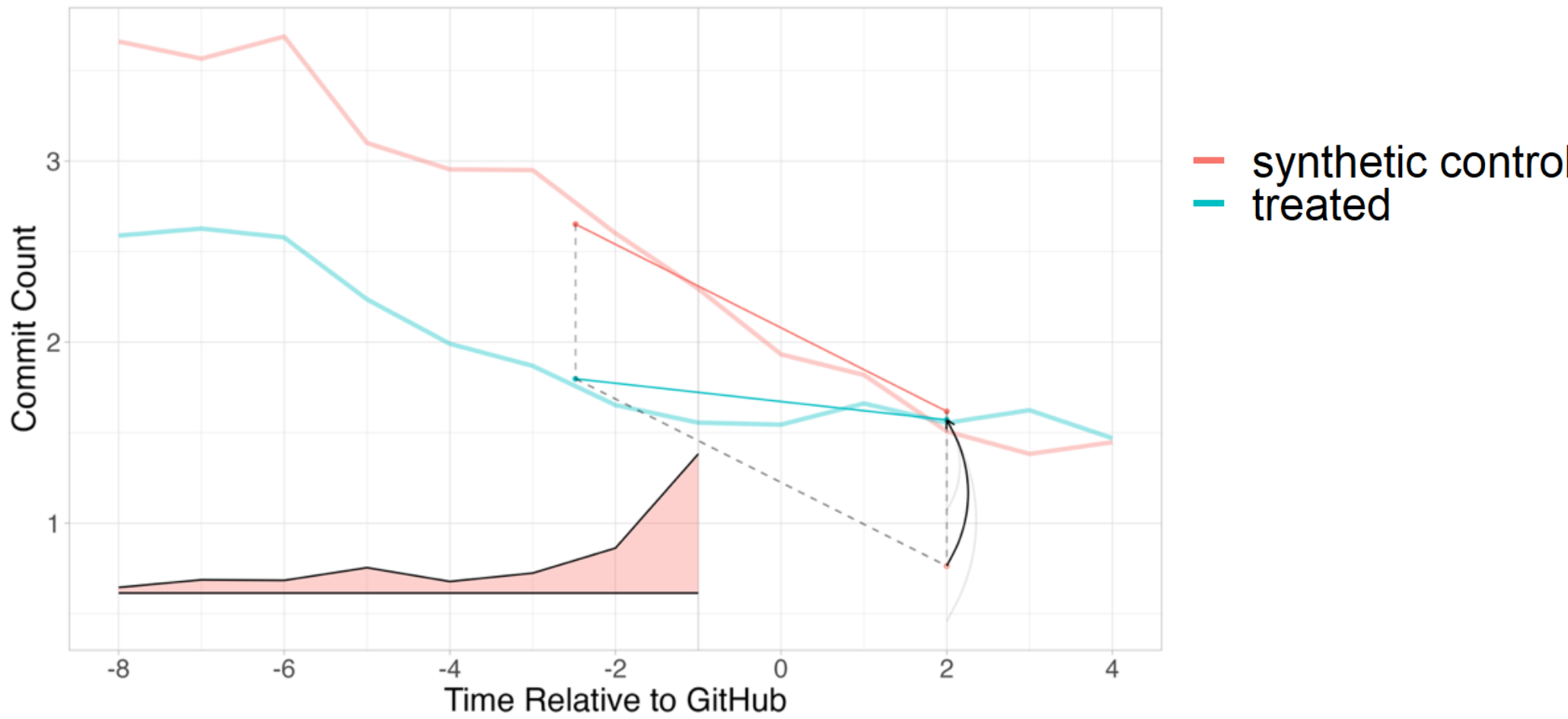

**Figure B6c.** The impact of Copilot on commits with new function definition (Combined: Python+Rust vs. R+Haskell)

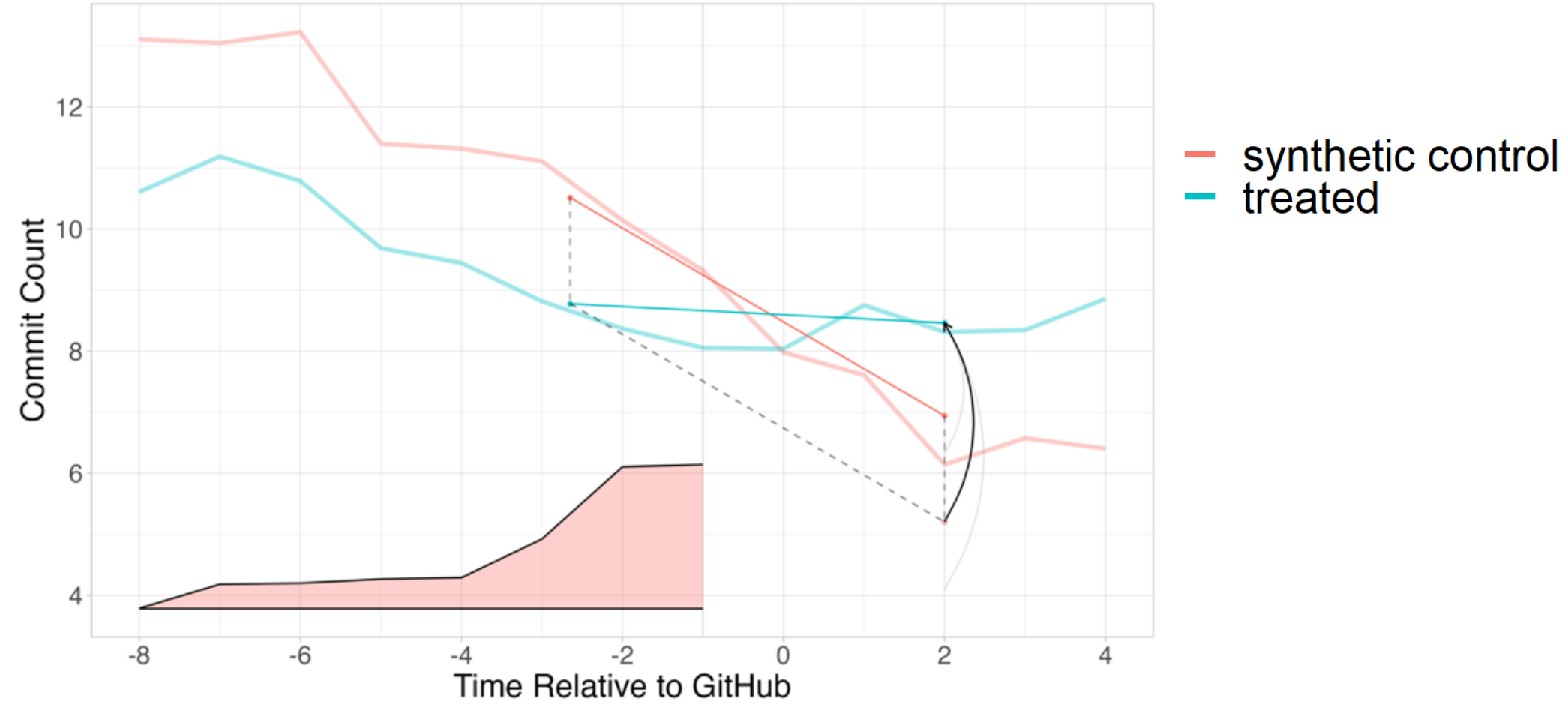

**Figure B6d.** The impact of Copilot on commits without new function definition (Combined: Python+Rust vs. R+Haskell)

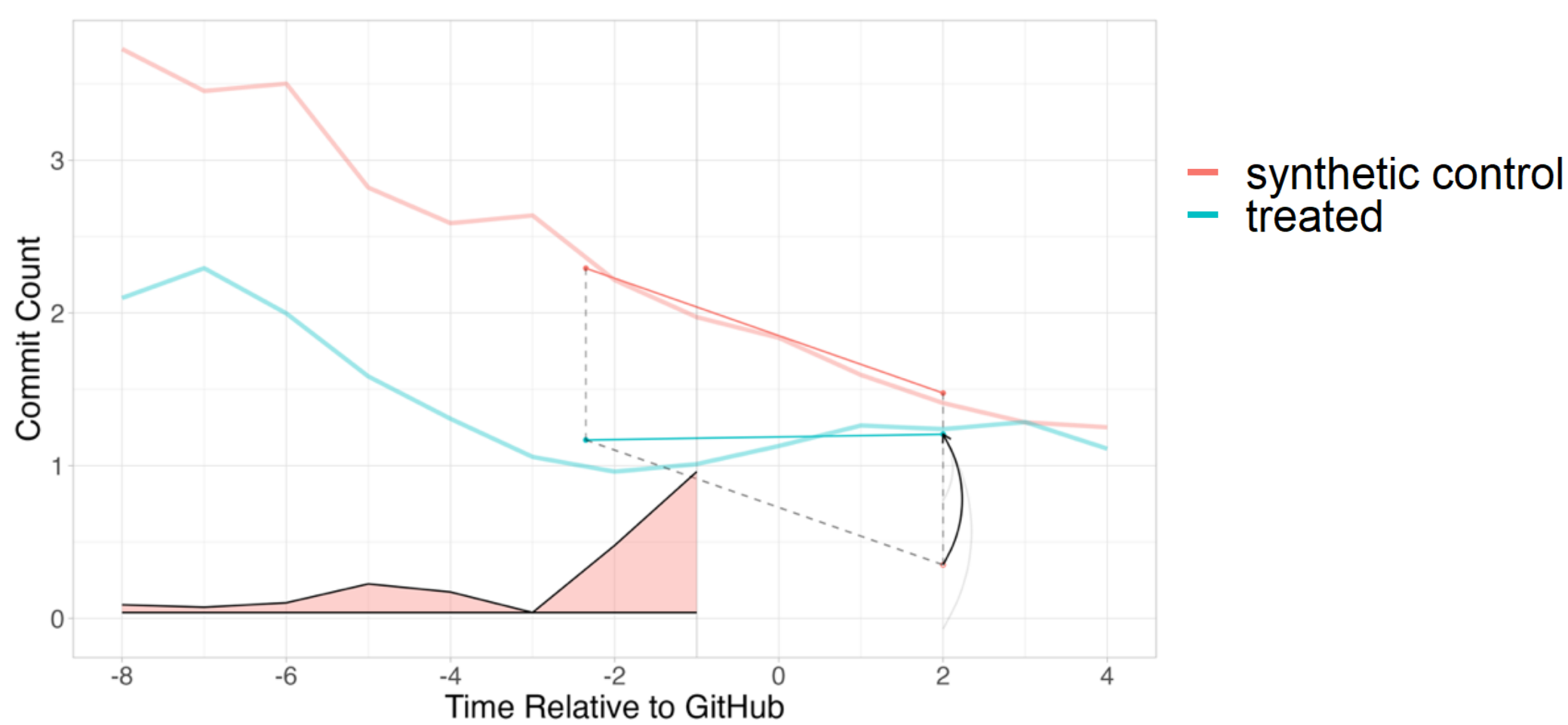

**Figure B7a.** The impact of Copilot on Code Development commits (Rust vs. Haskell)

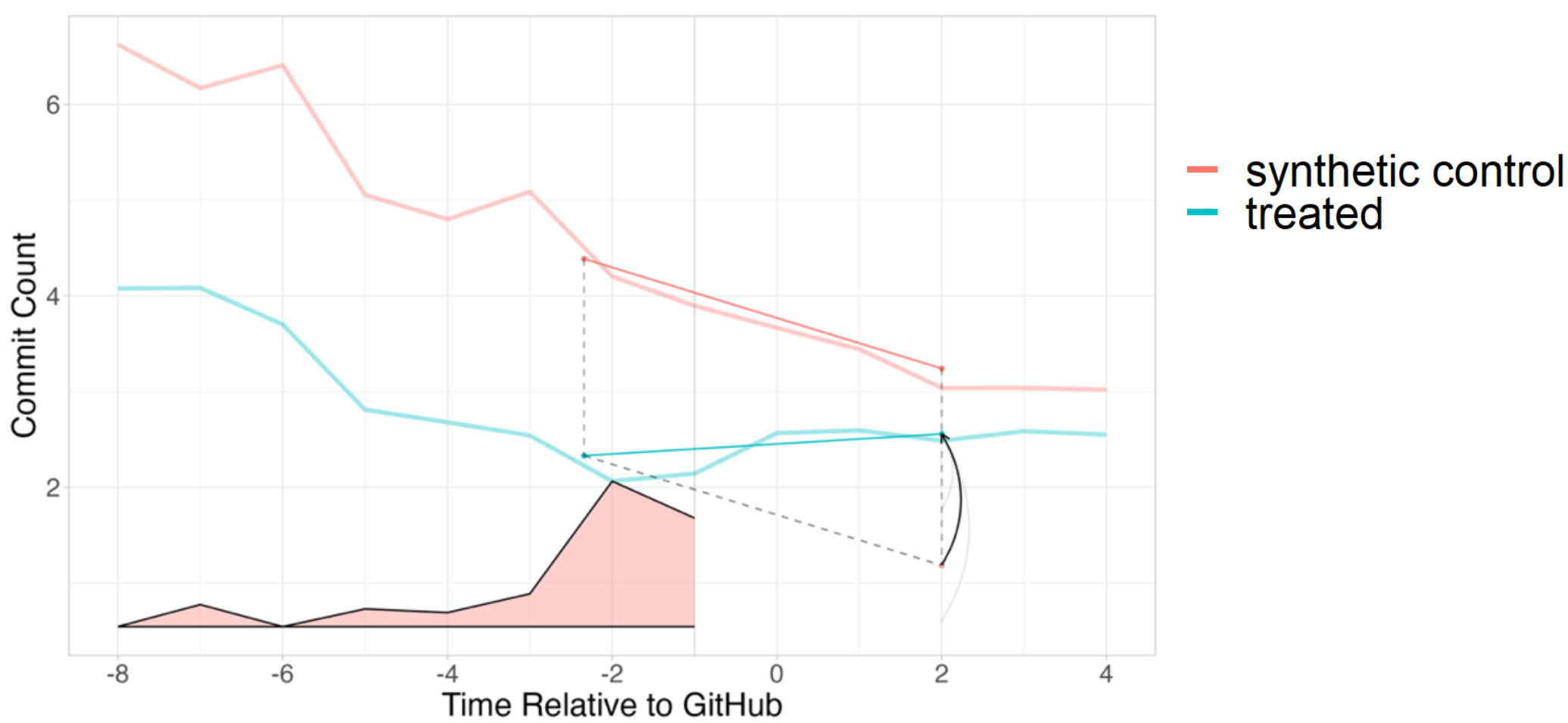

**Figure B7b.** The impact of Copilot on Maintenance commits (Rust vs. Haskell)

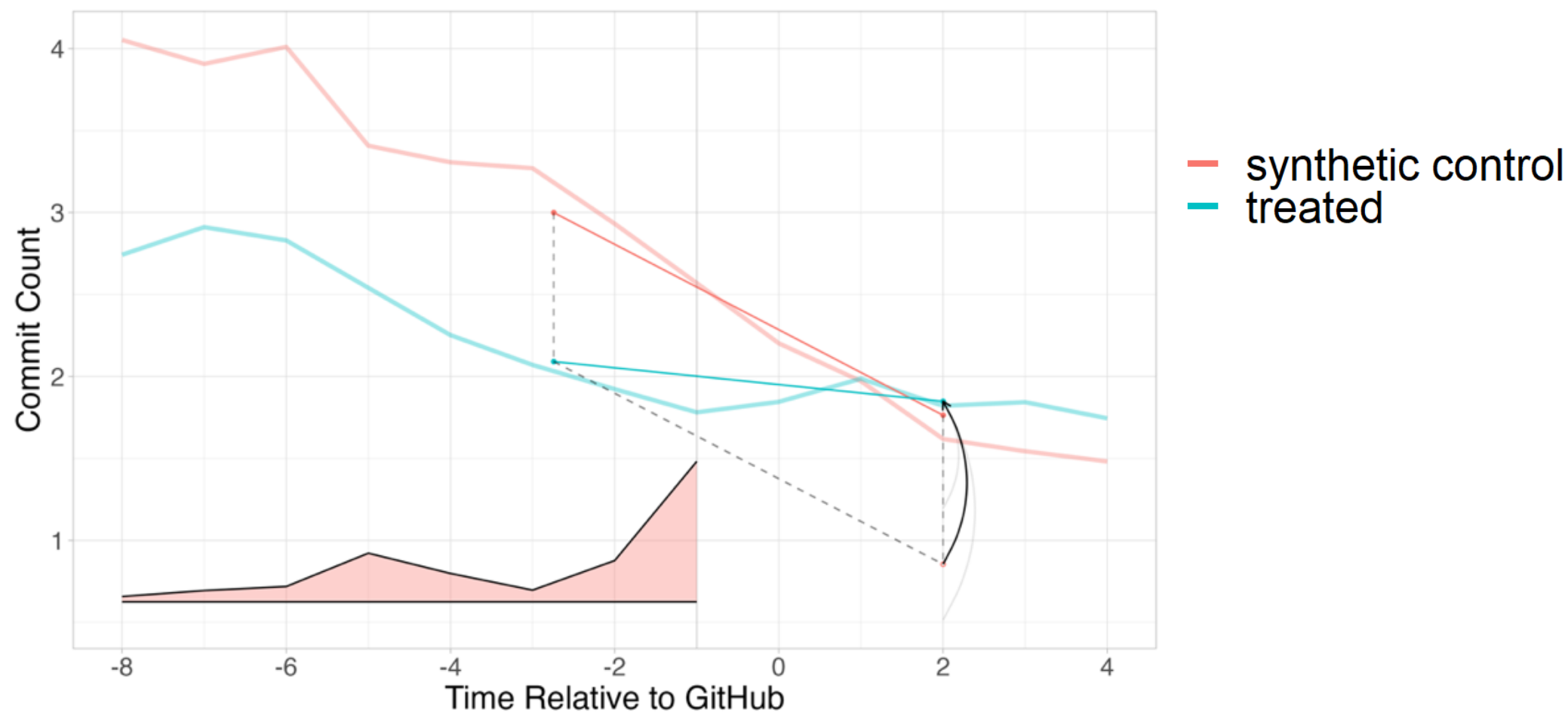

**Figure B7c.** The impact of Copilot on Code Development commits (Combined: Python+Rust vs. R+Haskell)

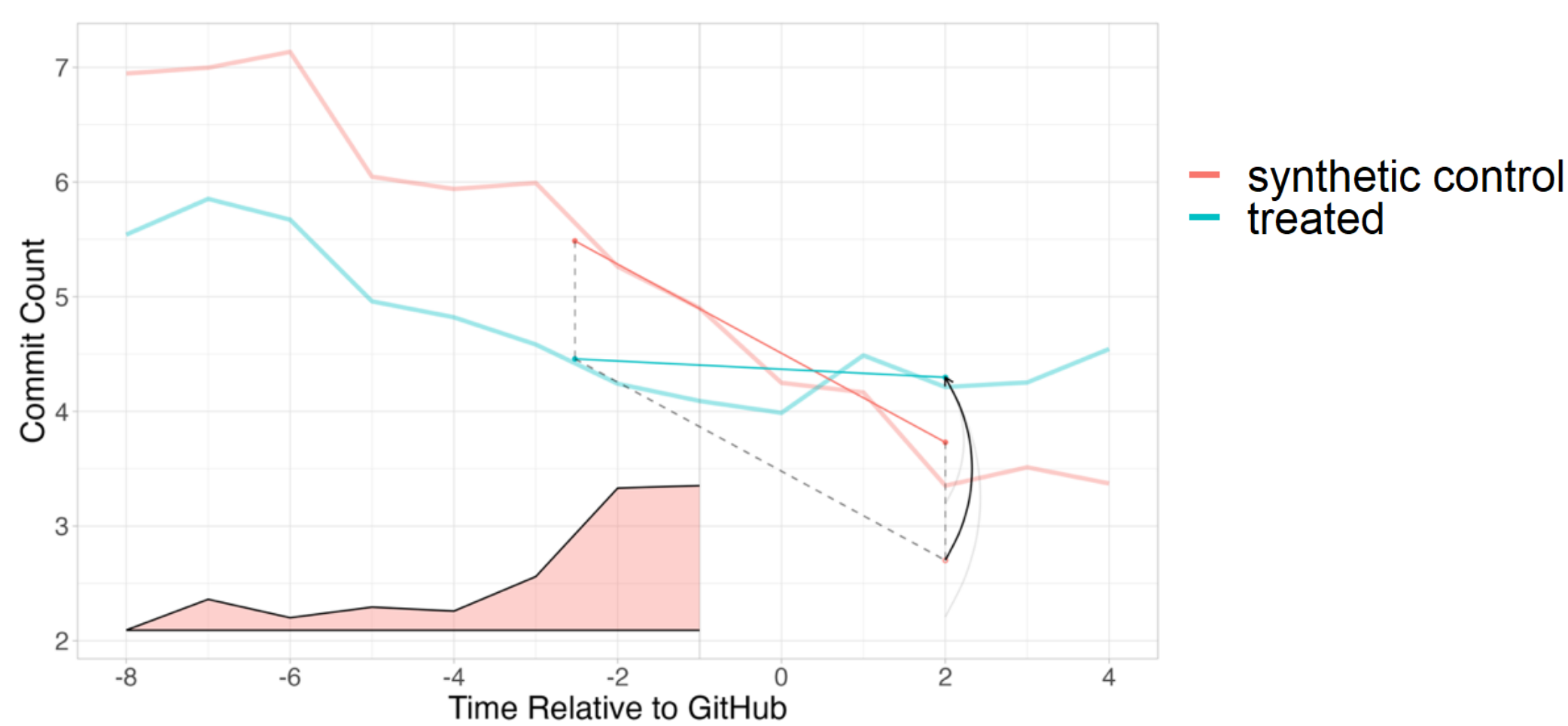

**Figure B7d.** The impact of Copilot on Maintenance commits (Combined: Python+Rust vs. R+Haskell)
Note: The x-axis indicates the period relative to the launch of GitHub Copilot (October 2021)

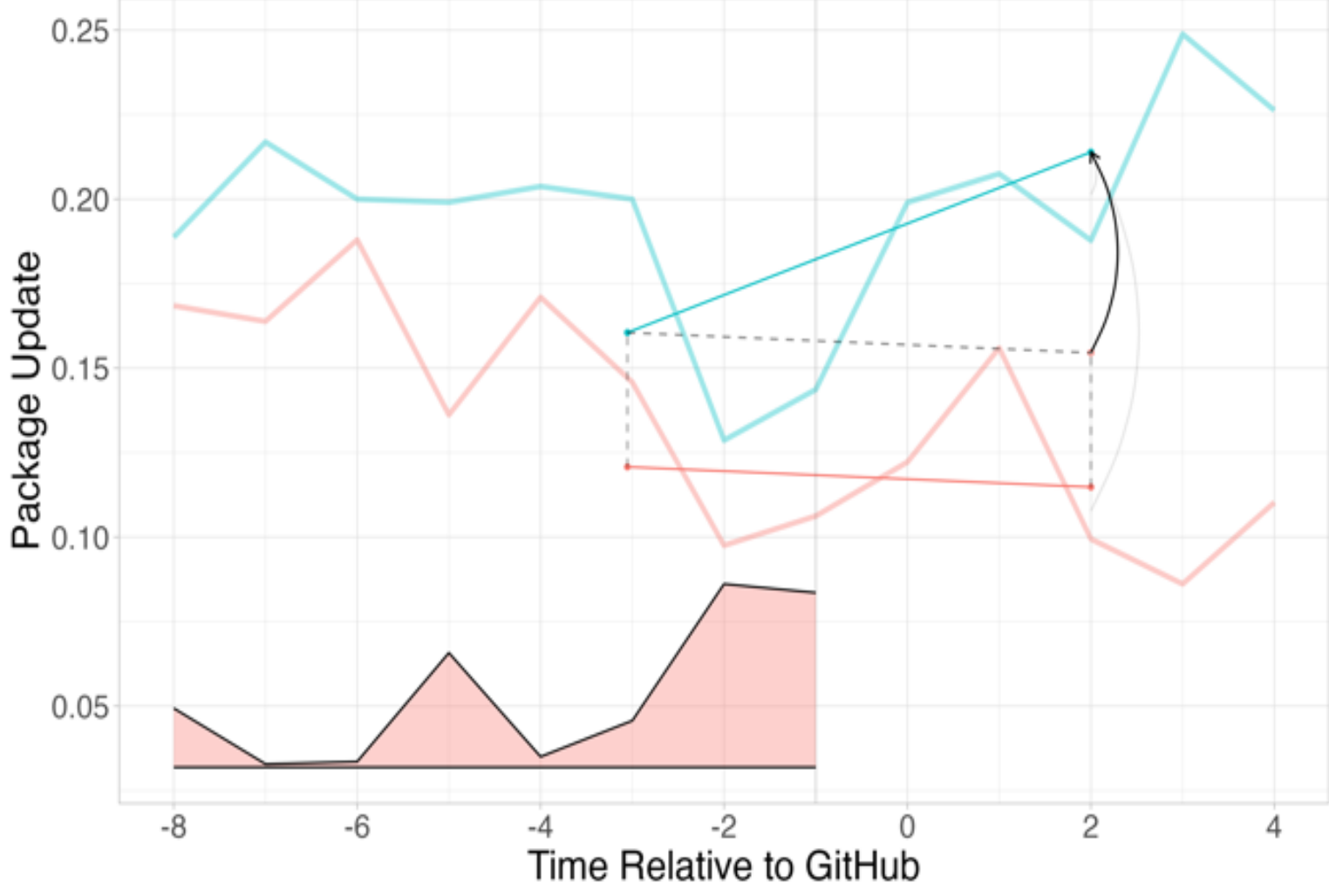

**Figure B8.** The impact of Copilot on New Version Release (Rust vs. Haskell)

# Appendix C

The primary task is to determine the category of commits made to GitHub using the commit "comments". The box below presents the annotation guidelines and the prompt that we used in employing LLMs for this classification task, developed using the guidelines by Cheng et al. (2026). The prompt is as follows:

You are an experienced programmer with extensive contributions to Python, R, Haskell and Rust projects. You will receive the name of a GitHub repository along with one of its commit comments. Your task is to meticulously analyze the commit comment and classify it into the most appropriate category out of the five categories based on its purpose or content. The five categories are:

1. Code Development: Includes 'Feature Addition' (adding new capabilities), 'Code Optimization' (enhancing performance, resource usage, efficiency, or setting and changing defaults), 'Algorithm implementation' (changing or adding new algorithms).
2. Maintenance: Encompasses 'Bug Fix' (correcting errors), 'Code Cleanup' (improving readability or structure without changing functionality), 'Dependency Management' (chores and grunt tasks for manual updating, manual version bump, or changing version requirements for dependencies), 'Refactoring' (altering code structure or organization to improve maintainability without changing behavior), 'Feature Removal or Deprecation' (taking out a feature from the codebase or marking it for future removal), 'Migration', 'Merge Operations'
3. Documentation and Style: Combines 'Documentation' (creating or updating documentation such as README, docstrings), 'Style Update' (modifying code styling for consistency or readability), new spoken language support and changelog.
4. Testing and Quality Assurance: Involves 'Testing' (adding or updating tests), 'Automatic Version Update and Numbers Management' (automatic bot triggered updating, managing and auto bumping version numbers or managing releases), 'automation' (including all automated updates, such as bot triggered version bumps) and logging (results log, test log etc.)
5. Other: Any changes that do not fit into the above categories (such as license and copyright updates, continuous integration continuous deployment, or mere repository operations without code or merging).

Additionally, assess the significance of the change as 'Minor', 'Medium', or 'Significant', and provide an estimated count of changes made. Include a concise justification for your classification.

Your response should be in a valid JSON format and strictly contain only the JSON, with all string values enclosed in double quotes. The response must include 'category' (a string), 'significance' (a string), 'count' (an integer, or a string if the count is unknown), and 'reason' (a string). For example:

```
{
  \"category\": \"Code Development\",
  \"significance\": \"Medium\",
  \"count\": 5,
  \"reason\": \"This commit adds a new feature improving data processing efficiency.\"
}
```

Remember to consider the context and implications of each commit comment while categorizing and assessing its impact. Do not include a double quote or commas inside the reason.

To choose the best model, we benchmarked some of their performances against human tagging. We first randomly selected 200 Python commit comments and 200 R commit comments and got them tagged by three expert human annotators. We then compared the outcome from five different LLMs (GPT 4o, GPT 4o mini, LLaMa 3.3 70B Q8.0, Deepseek R1 and Deepseek V3) with human tags and found that GPT4o and Deepseek V3 outperformed the other LLMs and were close enough. Table C1 shows the outcomes of the model and the Cohen's Kappa for the agreement between expert human annotators and the LLM predictions. The Cohen's Kappa is above 0.6 for gpt4o and deepseek models, which puts the agreement at the "substantial" level, as noted in extant research (e.g., Cohen 1960, McHugh 2012). Even though deepseek-chat was significantly cheaper than gpt-4o, we employed gpt-4o for time reasons for our annotation task. Table C2 presents the model details.

The total input tokens uploaded to the model for the entire process (1.1 million commits) were around 638 million tokens. The total output tokens generated by the model were approximately 60 million tokens.

**Table C1.** The accuracy of LLMs compared to human annotated ground truth.

| **Model** | **Agreement Rate** | **Cohen's Kappa** |
|---|---|---|
| gpt-4o-2024-08-06 | 79.50% | 0.72 |
| deepseek-reasoner (r1-2025-01-20) | 75.25% | 0.66 |
| deepseek-chat (v3-2024-12-26) | 76.50% | 0.67 |
| gpt4o-mini-2024-07-18 | 63.75% | 0.48 |
| Meta LLaMa 3.3 70B Q8.0 | 59.75% | 0.45 |

**Table C2.** The Model Parameters

| Model | endpoint | temperature | top_p | frequency_ penalty | presence_ penalty |
|---|---|---|---|---|---|
| gpt-4o-2024-08-06 | chat completion | 0.0 | 1.0 | 0.0 | 0.0 |

# Appendix D: Function Detection Methodology

We developed language-specific approaches to detect function definitions in code diffs across the four programming languages in our study. Here we outline the logic for each language.

### D.1 Python Function Detection

For Python, function detection focuses on identifying assignment operations that create functions:

- Check if the line contains the substring "`def`" (including the space)
- Count as a function if this substring is found
- Ensure that the line is not in the comment line (appears after #)

### D.2 R Function Detection

For R, function detection focuses on identifying assignment operations that create functions:

- Check if the line contains the substring "`function`" (including the space)
- Check if the line contains shorthand function notation introduced in R 4.1.0
- Count as a function if this substring is found
- Ensure that the line is not a comment line (appears after #)

### D.3 Rust Function Detection

For Rust, function detection focuses on identifying operations that create functions:

- Check if the line contains the substring "`fn`" (including the space)
- Count as a function if this substring is found
- Ensure that the line is not a comment line (appears after //)

### D.4 Haskell Function Detection

Haskell presented the greatest challenge because of its functional syntax. Our detection logic:

- Filter out non-function elements: comments, type signatures, module declarations, and data definitions.
- Check for function patterns with valid identifiers followed by parameters and an equals sign
- Handle nullary functions (functions without parameters).
- Account for pattern matching in function definitions.
- Exclude single-letter variables and Haskell keywords.
- Skip method implementations within instance blocks.